\documentclass[lettersize,journal]{IEEEtran}
\usepackage{amsmath,amssymb,amsfonts}
\usepackage{textcomp}

\usepackage{algorithm}
\usepackage[noend]{algpseudocode}

\usepackage{array}
\usepackage{textcomp}
\usepackage{stfloats}
\usepackage{float}
\usepackage{upgreek}
\usepackage{url}
\usepackage{verbatim}

\usepackage{graphicx}
\graphicspath{{figures/}}
\usepackage[labelformat=simple]{subcaption}

\usepackage[backend=biber, citestyle=numeric-comp, bibstyle=ieee]{biblatex}
\bibliography{refs}
\usepackage{comment}
\usepackage[prependcaption,textsize=footnotesize]{todonotes}
\usepackage{soul}
\usepackage{xcolor}
\usepackage{romannum}
\usepackage{pifont}

\usepackage{multirow}
\usepackage{booktabs}

\newenvironment{changed}
{}
{}

\usepackage{multirow}
\usepackage[normalem]{ulem}
\useunder{\uline}{\ul}{}

\begin{document}

\setlength{\abovedisplayskip}{2pt}
\setlength{\belowdisplayskip}{2pt}

\title{Mobility-Aware Resource Allocation for mmWave IAB Networks: A Multi-Agent Reinforcement Learning Approach}

\author{Bibo~Zhang
and~Ilario~Filippini,~\IEEEmembership{Senior Member,~IEEE}
\thanks{Bibo Zhang is with the Ocean College, Jiangsu University of Science and Technology, 212100 Zhenjiang, China (e-mail: bibo.zhang@just.edu.cn). Ilario Filippini is with Dipartimento di Elettronica, Informazione e Biongegneria, Politecnico di Milano, 20133 Milan,
Italy (e-mail: ilario.filippini@polimi.it).} 
}

\markboth{Submitted to IEEE/ACM Transactions on Networking}{}


\maketitle

\begin{abstract}
MmWaves have been envisioned as a promising direction to provide Gbps wireless access. However, they are susceptible to high path losses and blockages, which can only be partially mitigated by directional antennas. That makes mmWave networks coverage-limited, thus requiring dense deployments. Integrated access and backhaul (IAB) architectures have emerged as a cost-effective solution for network densification. Resource allocation in mmWave IAB networks must face big challenges originated by heavy temporal dynamics, such as intermittent links caused by user mobility and blockages from moving obstacles. This makes it extremely difficult to find optimal and adaptive solutions.
In this article, exploiting the distributed structure of the problem, we propose a Multi-Agent Reinforcement Learning (MARL) framework to optimize user throughput via flow routing and link scheduling in mmWave IAB networks characterized by mobile users and obstacles. The proposed approach implicitly captures the environment dynamics, coordinates the interference, and manages the buffer levels of IAB relay nodes. We design different MARL components, respectively for full-duplex and half-duplex networks. In addition, we propose an online training algorithm, which addresses the feasibility issues of practical systems, especially the communication and coordination among RL agents. Numerical results show the effectiveness of the proposed approach.
\end{abstract}


\vspace{-10pt}
\section{Introduction}
\label{sec:introduction}
\IEEEPARstart{T}{he} millimeter-wave (mmWave) bands have been considered by the 3rd Generation Partnership Project (3GPP) as one of the main reliefs to the explosive increase of the global mobile traffic, which is now posing big challenges to the access capacity provided by sub-6GHz communications.
Large bandwidths (several hundreds of MHz) and mainly-underutilized spectrum portions available at those high frequencies are the key enablers of a Gbps access throughput.
However, this potential comes at a cost of facing a harsh propagation environment characterized by very high path losses and weak, or even null, propagation through obstacles, including not only static buildings but also moving vehicles and pedestrians. While current antenna design technologies have shown to be effective in mitigating path losses through extremely directional arrays, there is very little they can do against random obstacle blockages. In practice, mmWave networks typically exhibit a coverage-limited behavior due to the presence of obstacles. Therefore, to guarantee a high-quality coverage, 5G mmWave access networks require base stations to be much more densely deployed than in traditional radio access networks. This may translate into high installation budgets for operators, which are mainly driven by costs to deploy wired (e.g., fiber) backhaul connections.

Aiming to provide a dense network deployment at minimal costs, 3GPP specifications have introduced in release 16 a new multi-hop wireless access architecture, named Integrated Access and Backhaul (IAB)\cite{3gpp15study}. The rationale is to place relay nodes, called IAB-nodes, in the service area of a mmWave base station (BS), called IAB-donor, and form a wireless multi-hop backhaul to forward data packets between the IAB-donor and user equipments (UEs). An example of such an IAB network scenario can be found in Fig.~\ref{fig:scenario}.
In recent years, regardless of technical challenges potentially posed, full-duplex IAB networks have been proposed in many works \cite{9770117,9363025} as a promising architecture for 5G NR paradigm\footnote{\begin{changed}3GPP has included the support for IAB simultaneous transmission and reception in Release 17 (3GPP TS 38.174 version 17.0.0 Release 17).\end{changed}}, which can considerably improve the spectral efficiency, compared with the commonly adopted half-duplex ones.
Self-backhauling is a peculiar aspect of IAB networks, where both radio access and backhaul links share the same radio resources and interfaces. Therefore, a proper radio resource allocation is essential to efficiently operate this network. In particular, since the adopted multiple access scheme is based on time-division multiple access (TDMA), the resource optimization must deal with routing paths and scheduling of directional transmissions along established links.

Studies on (joint) routing and scheduling in wireless multi-hop networks have appeared in the literature since early 2000s. It has been considered as a hard problem due to interference constraints and mainly solved resorting to optimization techniques that assume always-available links and static users \cite{yuan2018optimal}.
However, it is hard for these techniques to provide real-time solutions for dynamic mobile mmWave IAB networks.
Indeed, the optimal solution derived under ideal link conditions remarkably underperforms when facing the stochastic on-off link behavior caused by mobile users and the varying signal attenuation due to mobile obstacles.
In several cases, random link conditions can even eliminate all the advantages of a careful optimization.
The network could, in principle, be re-optimized periodically or every time it undergoes a change. However, it can induce huge computational costs and, most likely, not be practical, because a non-negligible amount of time is usually required to provide an optimal solution for a single network snapshot.
Therefore, flexible and adaptive solutions are required.

Given the above context, Reinforcement Learning (RL) techniques can play an important role thanks to their intrinsic adaptability to the environment behavior.
Indeed, an RL agent can automatically grasp relevant environment statistics by playing against the environment and eventually discover a strategy that can provide the best long-term reward.
However, several challenges need to be overcome if it is adopted.
First, access links are intermittently available due to UE mobility, which makes centralized single-agent RL (SARL) approaches infeasible. Indeed, their decision space is based on a set of potential concurrent transmissions (i.e., compatible links), which unfortunately changes as users randomly move around.
Second, randomly moving obstacles can dynamically cause different degrees of link attenuation, whose statistics must be learned.
Based on the above observations, random mobility patterns characterizing both UEs and obstacles typically generate local areas with local statistics that may vary across different network areas. To leverage such a scattered structure of the problem, multi-agent RL (MARL) techniques can be used to exploit multiple RL agents able to both make decisions based on local observations and coordinate with the other agents. This allows to split a single, complex, and high-dimensional problem -- which would otherwise be intractable -- into several, cooperative, and low-dimensional tasks.

In this article, we address the joint flow routing and link scheduling problem in mmWave IAB networks, coordinating both access and backhaul transmissions to maximize the downlink throughput perceived by UEs.
We provide an adaptive MARL-based framework that supports real-time operations and takes into account (1) physical constraints, including link interference, duplexing modes (i.e., full-duplex, half-duplex), hardware limitations, etc., (2) the amount of data cached in IAB-nodes' buffers, to avoid multi-hop flow starvation, (3) randomly moving 3D blocking obstacles, to reduce stochastic signal attenuation of mmWave communications to the full extent, and (4) randomly moving UEs, to dynamically adapt to changing access layouts.

Our contributions are summarized as follows:
\begin{itemize}
    \item By harnessing the distributed nature of the issue, we introduce an approach based on MARL that partitions a significantly intricate combinatorial resource allocation problem into numerous smaller tasks overseen by collaborative agents. This facilitates real-time execution of resource allocation operations that adjust to network conditions.
	\item The proposed approach can coordinate the interference among concurrent backhaul and access links, promptly monitor and refill IAB-node buffers to prevent downstream access transmission starvation, adjust transmission beams to serve mobile UEs, and adapt to intermittent blockages caused by randomly moving 3D obstacles.
	\item We develop distinct versions of our approach tailored for scenarios where all IAB-nodes operate either in full duplex (FD) or half duplex (HD) mode.
        \item We propose an online training framework that considers system-level aspects, particularly the challenges related to message exchange and coordination encountered in its practical application to mmWave IAB networks.
        \item We conduct extensive numerical experiments to evaluate the performance of the proposed approach and demonstrate how it can outperform the baselines.
\end{itemize}

\vspace{-10pt}
\section{Related Work}
\label{related_work}
In recent years, there have emerged many works on resource management for different types of networks. Table \ref{tab:related_work} summarizes these works, from aspects of network scenarios (w.r.t. network types, blockages, user mobility), resource management variables, objectives and techniques adopted.

Among all these works, traffic routing and transmission scheduling problems have been carefully studied
\cite{polese2018distributed, kilpi2017link,saad2019millimeter}.
However, only a few consider different link statuses (i.e., line-of-sight (LOS), non-LOS (NLOS), outages, etc.).
The works \cite{garcia2018delay,niu2019relay} find routing paths to bypass links interrupted by obstacles.
The work in \cite{he2015minimum} performs a slot-by-slot link scheduling to maximize the instantaneous throughput considering link blockage behaviors described by discrete-time Markov chains.
Authors in \cite{hu2017relay} deal with the relay selection and link scheduling problem to maximize the end-to-end throughput, using 3D models of buildings as primary blockage sources.
In \cite{yao2017outage}, heuristic algorithms for user scheduling and power allocation are proposed to reduce outage occurrences.
Our work differs from the above works in two aspects. First, we focus on randomly mobile 3D obstacles that can pose new challenges to the resource management problem, compared with static obstacles or stochastic-process-modeled blockages. Second, we aim to train intelligent transmission schemes that can make proper decisions by implicitly predicting the future moments.

User mobility in mmWave networks is mostly addressed in handover / user-cell association \cite{khan2019reinforcement,khosravi2020learning}, and only a few other types of resource management problems consider it.
The work in \cite{kim2019online} proposes a contextual multi-armed bandit algorithm to schedule transmissions to users with unknown positions.
In \cite{tang2020deep}, deep Q-network (DQN) is exploited to allocate capacity for up/down link transmissions in a 5G heterogeneous network with high-mobility.
Authors in \cite{pateromichelakis2018context} select routing paths based on the mobility and traffic conditions, then they perform a link-resource time sharing according to flow occupations.
A particle swarm optimization (PSO) algorithm is described in \cite{tafintsev2020aerial} to properly manage unmanned aerial vehicles (UAVs) in mmWave aerial access and backhaul networks to serve mobile users.
In \cite{shen2020mobility}, a band and beam allocation scheme for mmWave networks is presented, which considers massive multiple-input multiple-output (MIMO) systems and user trajectories.

However, none of these works deal with user mobility in coordinating concurrent backhaul and access transmission in mmWave IAB networks that are also characterized by link blockages, node buffers and different duplexing modes.

Recent years have witnessed a widespread utilization of RL techniques in mmWave networks.
A large part of the works in the literature deal with throughput maximization.
The work in \cite{lei2020deep} defines a spectrum allocation for IAB networks that maximizes the sum of log-rates through double DQN and actor critic techniques.
Authors in \cite{vu2018path} resort to regret RL and successive convex approximation to perform route selection and rate allocation, respectively.
Risk-sensitive RL is adopted in \cite{vu2018ultra} to control transmitter beamwidth and power so as to maximize data rate.
In \cite{zhang2021resource}, the authors propose a resource allocation framework based on advantage actor critic and column generation to maximize the throughput of static mmWave IAB networks.
Some of the other works investigate latency performances. Link scheduling approaches based on deep deterministic policy gradient (DDPG) \cite{gupta2020learning} and multi-armed bandit \cite{ortiz2019scaros} are proposed to minimize the end-to-end latency in mmWave backhaul networks.
Interference management and capacity issues have been faced as well. The work in \cite{feng2019dealing} allocates capacity between the core network and mmWave BSs to users subject to blockages. Authors in \cite{elsayed2020transfer} mitigate the inter-beam inter-cell interference through joint user-cell association and selection of number of beams.

The works \cite{vu2018path,zhang2021resource,gupta2020learning,ortiz2019scaros} study the path routing and link scheduling, however, \cite{zhang2021resource} did not consider UE mobility, while \cite{vu2018path,gupta2020learning} did not consider both link blockages and UE mobility.
Though \cite{ortiz2019scaros} considered both, it focuses on backhaul operations, emphasizing the load dynamics implicitly affected by user mobility statistics. Similarly, \cite{gupta2020learning} focuses on the backhaul part of an IAB network, while assuming static nodes and no blockages.

Finally, MARL emerges as a promising approach to cope with traffic signal control \cite{chu2019multi} and 
resource allocation \cite{kwon2019multi} in vehicular networks, user association \cite{sana2020multi} and handover management \cite{guo2020joint} in mmWave networks.
In \cite{kwon2019multi}, the authors propose a multi-agent deep deterministic policy gradient (MADDPG)-based approach for vehicles to select BSs to associate with and channels to communicate on, in order to maximize the system revenue.
The authors in \cite{sana2020multi} provide an MARL framework for static UEs to be associated with BSs, based on the hysteretic deep recurrent Q-network (HDRQN) algorithm.
In \cite{guo2020joint}, the authors jointly handle handover and power allocation to improve throughput and reduce handover frequency, by developing an MARL algorithm based on proximal policy optimization (PPO).
These works demonstrate good performance of MARL algorithms in making decisions in sophisticated systems, however, none of them deal with resource allocation problems considered in this work.

The above works set good examples of performing resource management in mmWave networks. However, there is still a lack of approaches that can maximize system throughput by adaptively performing flow allocation and link scheduling for both access and backhaul parts of mmWave IAB networks, whose dynamics are caused by both intermittent links and UE mobility. This work is motivated by such issues. 
In our previous work \cite{zhang2021mobility}, we have considered a simplified sector-based blockage model, a star backhaul topology, and only HD IAB-nodes. In this article, we extend it by considering a realistic link blockage model based on 3D mobile obstacles, a general tree-like backhaul topology, and both FD and HD working modes for IAB-nodes.

\begin{table*}[]
\centering
\caption{Summary of related work.}
\label{tab:related_work}
\resizebox{\textwidth}{!}{%
\begin{tabular}{|cc|ccccccc|cccc|}
\hline
\multicolumn{2}{|c|}{\multirow{2}{*}{}} &
  \multicolumn{7}{c|}{Resource management} &
  \multicolumn{4}{c|}{Objective} \\ \cline{3-13} 
\multicolumn{2}{|c|}{} &
  \multicolumn{1}{c|}{\begin{tabular}[c]{@{}c@{}}Routing/\\ path selection/\\ relay control\end{tabular}} &
  \multicolumn{1}{c|}{\begin{tabular}[c]{@{}c@{}}Transmission\\ scheduling\end{tabular}} &
  \multicolumn{1}{c|}{\begin{tabular}[c]{@{}c@{}}Capacity/\\ rate\\ allocation\end{tabular}} &
  \multicolumn{1}{c|}{\begin{tabular}[c]{@{}c@{}}Spectrum/\\ channel\\ allocation\end{tabular}} &
  \multicolumn{1}{c|}{\begin{tabular}[c]{@{}c@{}}Beam\\ allocation\end{tabular}} &
  \multicolumn{1}{c|}{\begin{tabular}[c]{@{}c@{}}Power\\ allocation\end{tabular}} &
  \begin{tabular}[c]{@{}c@{}}User-cell\\ association/\\ handover\end{tabular} &
  \multicolumn{1}{c|}{\begin{tabular}[c]{@{}c@{}}Throughput/\\ rate\\ maximization\end{tabular}} &
  \multicolumn{1}{c|}{\begin{tabular}[c]{@{}c@{}}Latency/\\ delay\\ minimization\end{tabular}} &
  \multicolumn{1}{c|}{\begin{tabular}[c]{@{}c@{}}Hops/\\ time length\\ minimization\end{tabular}} &
  \begin{tabular}[c]{@{}c@{}}Network\\ utility\\ maximization\end{tabular} \\ \hline
\multicolumn{2}{|c|}{IAB networks} &
  \multicolumn{1}{c|}{\begin{tabular}[c]{@{}c@{}}\cite{polese2018distributed,vu2018path,zhang2021resource}\\ \textbf{Ours}\end{tabular}} &
  \multicolumn{1}{c|}{\begin{tabular}[c]{@{}c@{}}\cite{zhang2021resource}\\ \textbf{Ours}\end{tabular}} &
  \multicolumn{1}{c|}{\cite{vu2018path}} &
  \multicolumn{1}{c|}{\cite{lei2020deep}} &
  \multicolumn{1}{c|}{} &
  \multicolumn{1}{c|}{} &
   &
  \multicolumn{1}{c|}{\begin{tabular}[c]{@{}c@{}}\cite{polese2018distributed,lei2020deep,zhang2021resource}\\ \textbf{Ours}\end{tabular}} &
  \multicolumn{1}{c|}{} &
  \multicolumn{1}{c|}{\cite{polese2018distributed}} &
  \cite{vu2018path} \\ \hline
\multicolumn{2}{|c|}{\begin{tabular}[c]{@{}c@{}}mmWave \\ backhaul networks\end{tabular}} &
  \multicolumn{1}{c|}{\cite{garcia2018delay,niu2019relay,hu2017relay,ortiz2019scaros}} &
  \multicolumn{1}{c|}{\begin{tabular}[c]{@{}c@{}}\cite{kilpi2017link,saad2019millimeter,garcia2018delay,niu2019relay,hu2017relay}\\ \cite{ortiz2019scaros,gupta2020learning}\end{tabular}} &
  \multicolumn{1}{c|}{} &
  \multicolumn{1}{c|}{} &
  \multicolumn{1}{c|}{} &
  \multicolumn{1}{c|}{} &
   &
  \multicolumn{1}{c|}{\cite{saad2019millimeter,garcia2018delay,niu2019relay, hu2017relay}} &
  \multicolumn{1}{c|}{\cite{kilpi2017link,garcia2018delay,gupta2020learning,ortiz2019scaros}} &
  \multicolumn{1}{c|}{\cite{kilpi2017link}} &
   \\ \hline
\multicolumn{2}{|c|}{Other networks} &
  \multicolumn{1}{c|}{\cite{he2015minimum,pateromichelakis2018context,tafintsev2020aerial}} &
  \multicolumn{1}{c|}{\begin{tabular}[c]{@{}c@{}}\cite{he2015minimum,yao2017outage}\\ \cite{kim2019online,pateromichelakis2018context}\end{tabular}} &
  \multicolumn{1}{c|}{\cite{tang2020deep,feng2019dealing}} &
  \multicolumn{1}{c|}{\cite{kwon2019multi}} &
  \multicolumn{1}{c|}{\begin{tabular}[c]{@{}c@{}}\cite{khosravi2020learning,shen2020mobility}\\ \cite{vu2018ultra,elsayed2020transfer}\end{tabular}} &
  \multicolumn{1}{c|}{\begin{tabular}[c]{@{}c@{}}\cite{yao2017outage,vu2018ultra}\\ \cite{guo2020joint}\end{tabular}} &
  \begin{tabular}[c]{@{}c@{}}\cite{khan2019reinforcement,sana2020multi,elsayed2020transfer}\\ \cite{kwon2019multi,khosravi2020learning,guo2020joint}\end{tabular} &
  \multicolumn{1}{c|}{\begin{tabular}[c]{@{}c@{}}\cite{he2015minimum,khan2019reinforcement,khosravi2020learning}\\ \cite{kim2019online,tang2020deep,pateromichelakis2018context,tafintsev2020aerial,shen2020mobility,feng2019dealing}\\ \cite{elsayed2020transfer,sana2020multi,guo2020joint}\end{tabular}} &
  \multicolumn{1}{c|}{} &
  \multicolumn{1}{c|}{} &
  \cite{vu2018ultra,kwon2019multi} \\ \hline
\multicolumn{2}{|c|}{Blockages} &
  \multicolumn{1}{c|}{\begin{tabular}[c]{@{}c@{}}\cite{garcia2018delay,niu2019relay,he2015minimum,hu2017relay,zhang2021resource,ortiz2019scaros}\\ \textbf{Ours}\end{tabular}} &
  \multicolumn{1}{c|}{\begin{tabular}[c]{@{}c@{}}\cite{garcia2018delay,niu2019relay,he2015minimum,hu2017relay,yao2017outage}\\ \cite{zhang2021resource,ortiz2019scaros}\\ \textbf{Ours}\end{tabular}} &
  \multicolumn{1}{c|}{\cite{feng2019dealing}} &
  \multicolumn{1}{c|}{} &
  \multicolumn{1}{c|}{\cite{shen2020mobility,vu2018ultra}} &
  \multicolumn{1}{c|}{\cite{yao2017outage,vu2018ultra}} &
   &
  \multicolumn{1}{c|}{\begin{tabular}[c]{@{}c@{}}\cite{garcia2018delay,niu2019relay,he2015minimum,hu2017relay,shen2020mobility,zhang2021resource}\\ \cite{feng2019dealing}\\ \textbf{Ours}\end{tabular}} &
  \multicolumn{1}{c|}{\cite{garcia2018delay,ortiz2019scaros}} &
  \multicolumn{1}{c|}{\cite{yao2017outage,he2015minimum}} &
  \cite{vu2018ultra} \\ \hline
\multicolumn{2}{|c|}{UE mobility} &
  \multicolumn{1}{c|}{\begin{tabular}[c]{@{}c@{}}\cite{pateromichelakis2018context,tafintsev2020aerial,ortiz2019scaros}\\ \textbf{Ours}\end{tabular}} &
  \multicolumn{1}{c|}{\begin{tabular}[c]{@{}c@{}}\cite{yao2017outage,kim2019online}\\ \cite{pateromichelakis2018context,ortiz2019scaros}\\ \textbf{Ours}\end{tabular}} &
  \multicolumn{1}{c|}{\cite{tang2020deep}} &
  \multicolumn{1}{c|}{\cite{lei2020deep,kwon2019multi}} &
  \multicolumn{1}{c|}{\begin{tabular}[c]{@{}c@{}}\cite{khosravi2020learning,shen2020mobility}\\ \cite{elsayed2020transfer}\end{tabular}} &
  \multicolumn{1}{c|}{\cite{yao2017outage,guo2020joint}} &
  \begin{tabular}[c]{@{}c@{}}\cite{khan2019reinforcement,elsayed2020transfer,kwon2019multi}\\ \cite{khosravi2020learning,guo2020joint}\end{tabular} &
  \multicolumn{1}{c|}{\begin{tabular}[c]{@{}c@{}}\cite{khan2019reinforcement,khosravi2020learning,kim2019online,tang2020deep,pateromichelakis2018context,tafintsev2020aerial,shen2020mobility,lei2020deep}\\ \cite{elsayed2020transfer,guo2020joint}\\ \textbf{Ours}\end{tabular}} &
  \multicolumn{1}{c|}{\cite{ortiz2019scaros}} &
  \multicolumn{1}{c|}{\cite{yao2017outage}} &
  \cite{kwon2019multi} \\ \hline
\multicolumn{1}{|c|}{\multirow{5}{*}{SARL}} &
  MAB &
  \multicolumn{1}{c|}{\cite{ortiz2019scaros}} &
  \multicolumn{1}{c|}{\cite{kim2019online,ortiz2019scaros}} &
  \multicolumn{1}{c|}{} &
  \multicolumn{1}{c|}{} &
  \multicolumn{1}{c|}{} &
  \multicolumn{1}{c|}{} &
   &
  \multicolumn{1}{c|}{\cite{kim2019online}} &
  \multicolumn{1}{c|}{\cite{ortiz2019scaros}} &
  \multicolumn{1}{c|}{} &
   \\ \cline{2-13} 
\multicolumn{1}{|c|}{} &
  QL-based &
  \multicolumn{1}{c|}{} &
  \multicolumn{1}{c|}{} &
  \multicolumn{1}{c|}{} &
  \multicolumn{1}{c|}{} &
  \multicolumn{1}{c|}{\cite{khosravi2020learning,elsayed2020transfer}} &
  \multicolumn{1}{c|}{} &
  \cite{elsayed2020transfer,khosravi2020learning} &
  \multicolumn{1}{c|}{\cite{khosravi2020learning,elsayed2020transfer}} &
  \multicolumn{1}{c|}{} &
  \multicolumn{1}{c|}{} &
   \\ \cline{2-13} 
\multicolumn{1}{|c|}{} &
  DQN-based &
  \multicolumn{1}{c|}{} &
  \multicolumn{1}{c|}{} &
  \multicolumn{1}{c|}{\cite{tang2020deep,feng2019dealing}} &
  \multicolumn{1}{c|}{\cite{lei2020deep}} &
  \multicolumn{1}{c|}{} &
  \multicolumn{1}{c|}{} &
   &
  \multicolumn{1}{c|}{\cite{tang2020deep,lei2020deep,feng2019dealing}} &
  \multicolumn{1}{c|}{} &
  \multicolumn{1}{c|}{} &
   \\ \cline{2-13} 
\multicolumn{1}{|c|}{} &
  AC-based &
  \multicolumn{1}{c|}{\cite{zhang2021resource}} &
  \multicolumn{1}{c|}{\cite{zhang2021resource,gupta2020learning}} &
  \multicolumn{1}{c|}{} &
  \multicolumn{1}{c|}{\cite{lei2020deep}} &
  \multicolumn{1}{c|}{} &
  \multicolumn{1}{c|}{} &
   &
  \multicolumn{1}{c|}{\cite{lei2020deep,zhang2021resource}} &
  \multicolumn{1}{c|}{\cite{gupta2020learning}} &
  \multicolumn{1}{c|}{} &
   \\ \cline{2-13} 
\multicolumn{1}{|c|}{} &
  Others &
  \multicolumn{1}{c|}{\cite{vu2018path}} &
  \multicolumn{1}{c|}{} &
  \multicolumn{1}{c|}{\cite{vu2018path}} &
  \multicolumn{1}{c|}{} &
  \multicolumn{1}{c|}{\cite{vu2018ultra}} &
  \multicolumn{1}{c|}{\cite{vu2018ultra}} &
   &
  \multicolumn{1}{c|}{} &
  \multicolumn{1}{c|}{} &
  \multicolumn{1}{c|}{} &
  \cite{vu2018path,vu2018ultra} \\ \hline
\multicolumn{2}{|c|}{MARL} &
  \multicolumn{1}{c|}{\textbf{Ours}} &
  \multicolumn{1}{c|}{\textbf{Ours}} &
  \multicolumn{1}{c|}{} &
  \multicolumn{1}{c|}{\cite{kwon2019multi}} &
  \multicolumn{1}{c|}{} &
  \multicolumn{1}{c|}{\cite{guo2020joint}} &
  \cite{khan2019reinforcement,kwon2019multi,guo2020joint,sana2020multi} &
  \multicolumn{1}{c|}{\begin{tabular}[c]{@{}c@{}}\cite{khan2019reinforcement,guo2020joint,sana2020multi}\\ \textbf{Ours}\end{tabular}} &
  \multicolumn{1}{c|}{} &
  \multicolumn{1}{c|}{} &
  \cite{kwon2019multi} \\ \hline
\end{tabular}%
}
\end{table*}

\vspace{-10pt}
\section{System Model}
\label{system_model}
\begin{figure}
	\centering
	\includegraphics[width=0.4\textwidth]{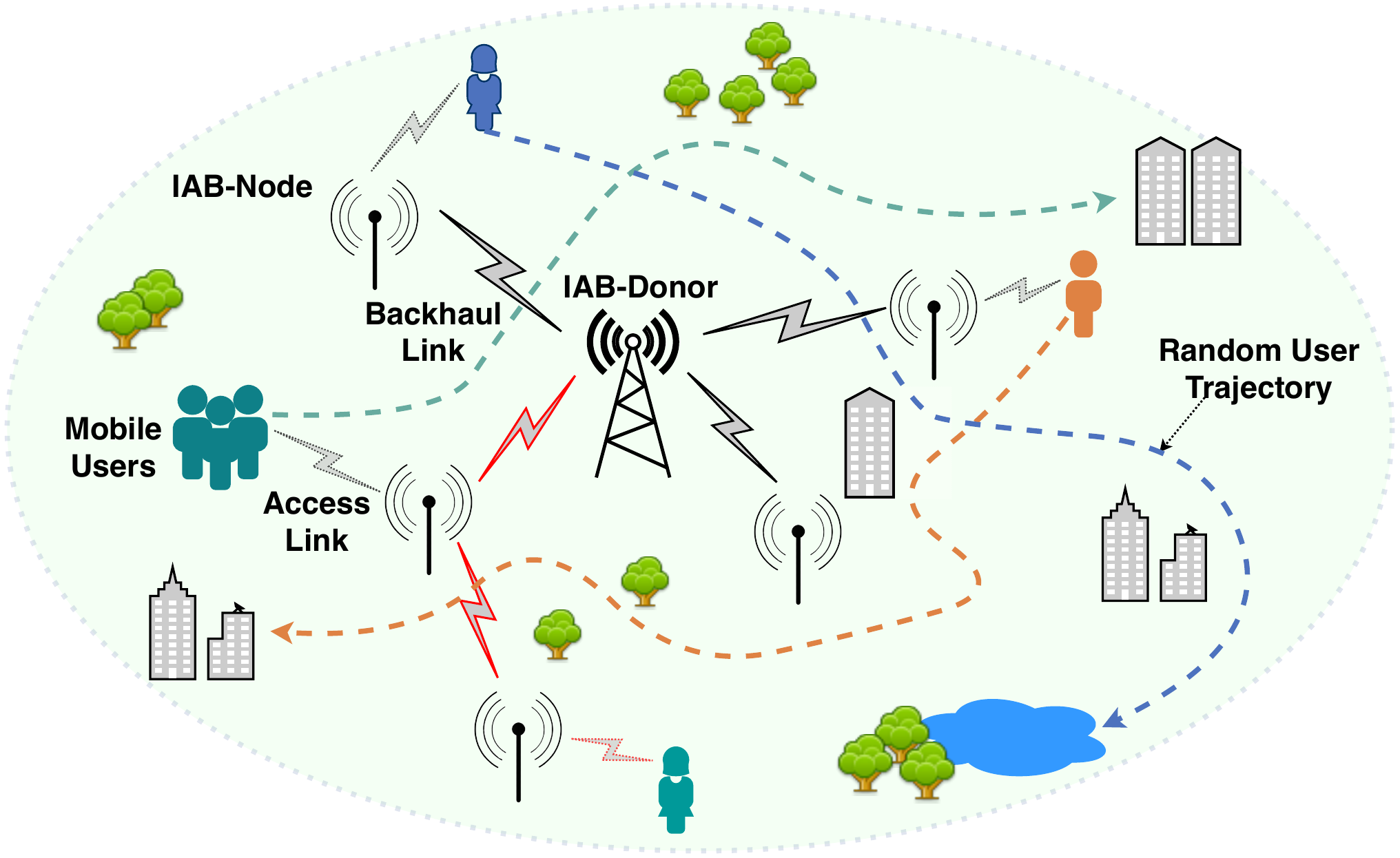}
	\caption{\small An example of IAB network scenario with mobile users. The dashed arrows represent user trajectories.}
	\label{fig:scenario}
	\vspace{-6mm}
\end{figure}

We consider a mmWave IAB network that consists of a mmWave BS,\textit{ IAB-donor}, connected to the core network through a wired connection, and a set of small mmWave BSs,\textit{ IAB-nodes}, wirelessly backhauled to the IAB-donor using mmWave frequencies.
Mobile UEs that move in the service area according to the well-known Random Waypoint model \cite{bai2004survey} get access to the network via either direct mmWave links connected with the IAB-donor or multiple hops through IAB-nodes. An example of IAB network scenario is depicted in Fig. \ref{fig:scenario}.
Backhaul links are established either between the IAB-donor and an IAB-node or between two IAB-nodes, while access links connect IAB-donor/IAB-nodes to UEs. Both backhaul and access transmissions share the same mmWave frequency band (i.e., in-band backhaul).

This scenario can be represented as a graph $\mathcal{G(\mathcal{V}, \mathcal{E})}$ where the node set $\mathcal{V}$ consists of an IAB-donor, IAB-nodes and UEs, and the edge set $\mathcal{E}$ includes all the potential links among the nodes. $\mathcal{V}$ can be further divided into the set of IAB-nodes $\mathcal{R}$ and the set of UEs $\mathcal{U}$. If not differently specified, the IAB-donor is identified as a special IAB-node.
We consider a tree topology for the backhaul network, where IAB-nodes are connected to the IAB-donor either directly or via multiple hops, as indicated in 3GPP specifications for IAB networks \cite{3gpp14studyPhysical}, which assume no more than 10 IAB-nodes organized in simple topologies.

\subsection{Channel Model}
\label{blockage_mobility_model}

We adopt typical path loss and antenna pattern models \cite{maltsev2014d5} for mmWave communications. The path loss model, considering both the line-of-sight (LOS) component and close reflections from the ground and other objects, is defined as \cite{maltsev2014d5} Eq. 5-37:
\begin{equation}
	PL_{dB} = \alpha + k\cdot10\cdot \text{Log}\left(\frac{d}{d_0}\right),
	\label{pathloss}
\end{equation}
where $\alpha=82.02dB$ and $d_0=5m$. $d$ is the path length. The propagation factor $k$ is 2.36 if $d > d_0$ and 2 otherwise.
The antenna gain is modeled with a Gaussian main lobe profile \cite{maltsev2014d5} Eq. 5-32:
\begin{equation}
	G_{dB}(\phi, \theta) = 10\cdot\text{Log}(G_0) - 12\cdot\frac{\phi^2}{\phi^2_{-3dB}} - 12\cdot\frac{\beta^2}{\beta^2_{-3dB}},
	\label{antenna_gain}
\end{equation}
\begin{equation}
	G_0 = \frac{16\pi}{6.76\cdot\phi_{-3dB}\cdot\beta_{-3dB}}.
 \label{base_antenna_gain}
\end{equation}
$\phi_{-3dB}$ and $\beta_{-3dB}$ are respectively the elevation and azimuth half power beam widths (HPBWs). The $\phi$ and $\beta$ are the elevation and azimuth angle offsets between the main lobe direction and the direction to the considered transmitter/receiver.
A visible definition of $\phi_{-3dB}$, $\beta_{-3dB}$, $\phi$ and $\beta$ can be found in Figs. 5-18, 5-19 and 5-20 in \cite{maltsev2014d5}.

Non-line-of-sight (NLOS) conditions are mainly caused by blocking obstacles. IAB-nodes are expected to be installed at relatively high places (e.g., street lights, roof tops, etc.) to improve visibility and avoid tampering (e.g., the height of IAB-nodes is typically assumed to be about $6$m, which is hard to be reached even for a double-decker bus), thus we expect that mobile obstacles can rarely affect backhaul links. Nevertheless, there can still be some static obstacles, such as buildings, causing severe blockages to backhaul links. However, they can be effectively avoided in the network planning stage by considering radio coverage.
In contrast, access links are exposed to more recurrent blockages caused by nomadic obstacles (e.g., pedestrian, transportation traffic, etc.). Based on these observations, we realistically apply random and mobile obstacle blockages only to access links.

\begin{figure}
	\centering
 \includegraphics[width=0.45\textwidth]{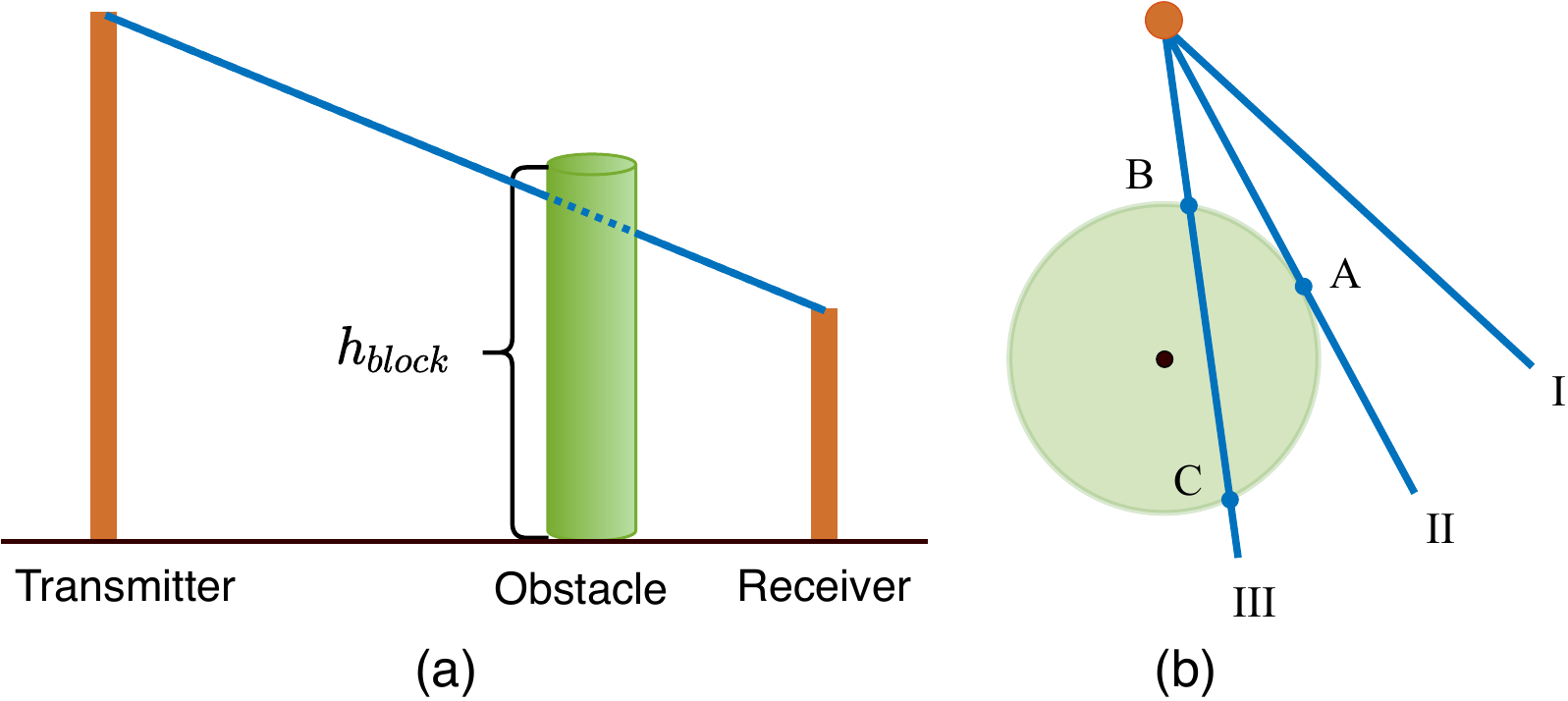}
	\caption{\small Blockage model considering 3D obstacles.}
	\label{fig:3d_blockage}
	\vspace{-4mm}
\end{figure}

We consider 3D mobile obstacles \cite{gapeyenko2017temporal} that move
according to Random Waypoint model \cite{bai2004survey}.
Fig. \ref{fig:3d_blockage}(a) shows an example  of how the blockage between a transmitter and a receiver occurs.
An obstacle is modeled as a cylinder standing in the LOS path between the transmitter and the receiver.
Fig. \ref{fig:3d_blockage}(b) shows the corresponding top view where the intersection points between the LOS path and the blocking cylinder are identified as $A$, $B$ and $C$, having respectively heights of $h_A$, $h_B$ and $h_C$. To determine whether a link blockage occurs, the following cases, represented in Fig. \ref{fig:3d_blockage}(b), are examined:
\begin{itemize}
	\item Case \Romannum{1}: when the LOS path is neither a secant nor a tangent of the blocker's cross-section, there is no blockage in the transmission.
	\item Case \Romannum{2}: when the LOS path is a tangent of the blocker's cross-section, if $h_A \leq h_{block}$, the blockage occurs; otherwise, no blockage occurs.
	\item Case \Romannum{3}: when the LOS path is a secant of the blocker's cross-section. A blockage occurs only if $h_B \leq h_{block}$ or $h_C \leq h_{block}$.
\end{itemize}
Every time an obstacle interrupts a transmission, the link blockage occurs according to a knife-edge diffraction model indicated in the 3GPP specification \cite{3gpp15studyChannel}, where the additional attenuation caused by each obstacle is applied to the path loss model \eqref{pathloss}.
We would like to remark that the above models are just reasonable choices to obtain realistic scenarios, and thus meaningful numerical results. Indeed, the approach we propose can be applied to other specific path loss, antenna pattern and blockage models as well.

\subsection{Network Designs}

\begin{figure*}[ht]
	\centering
	\includegraphics[width=0.8\linewidth]{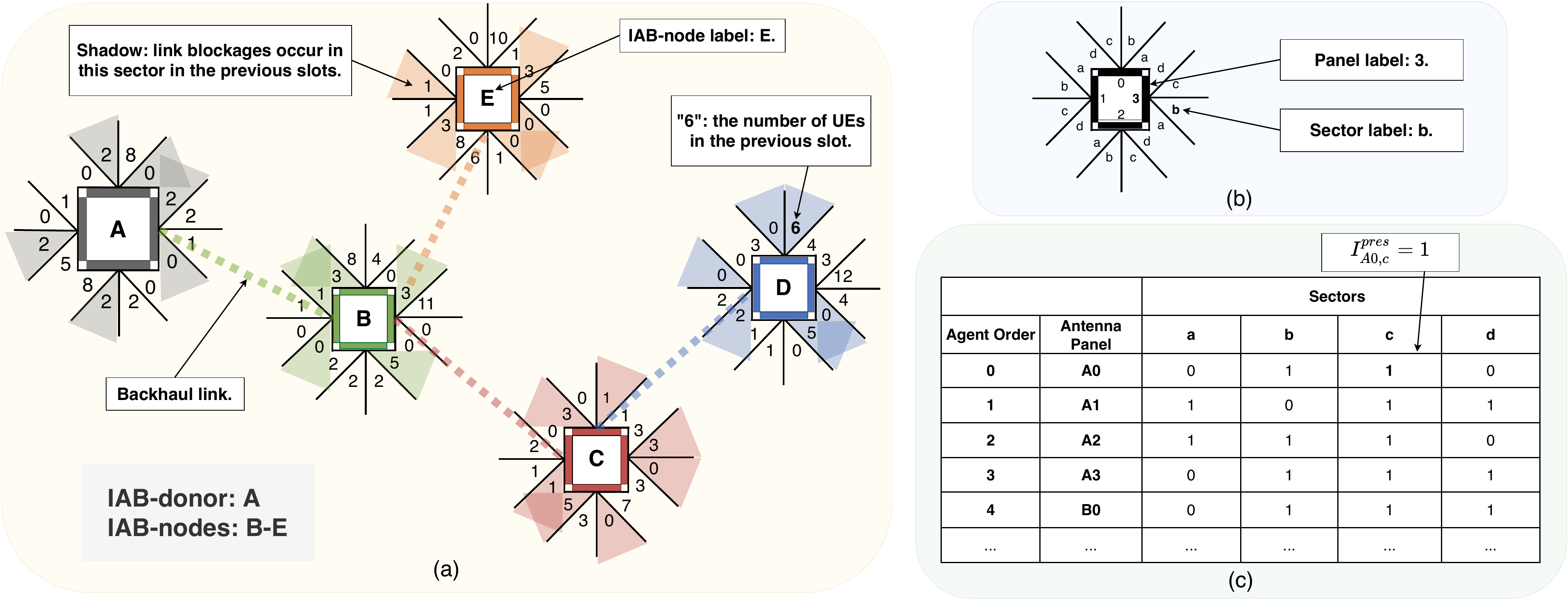}
	\caption{\footnotesize A top-view example of IAB network scenario and its corresponding RL formulation. \textbf{(a)} shows an IAB network with 1 IAB-donor and 4 IAB-nodes. Backhaul links (dashed lines) form a tree topology and the number of covered UEs is reported in each sector. A sector is shadowed if blockages have been detected in it in the previous slots. \textbf{(b)} details an IAB-node equipped with $N_p=4$ Tx array panels (1, 2, 3, 4), each of which manages $N_s=4$ sectors (a, b, c, d). \textbf{(c)} includes a table recording the binary UE presence information for the sectors shown in (a).
	}
	\label{fig:principle}
    \vspace{-4mm}
\end{figure*}

In this work, we aim to investigate downlink traffic transfer from the IAB-donor to UEs, in order to provide maximum UE throughput in IAB networks with the following designs. 
All the IAB-nodes together operate either in HD mode (i.e., either receiving or transmitting data in the same slot) or in FD mode (i.e., able to simultaneously receiving and transmitting data). An HD IAB-node (or IAB-donor) is equipped with $N_p$ side-by-side array panels that simultaneously manage transmission (Tx) and reception (Rx), while a FD IAB-node is equipped with $N_p$ side-by-side pairs of separate Tx and Rx array panels\footnote{In the experiments carried out in Sec.~\ref{experiments}, we assume ideal self-interference cancellation and isolation for FD IAB-nodes, however, the proposed approach can also be applied in scenarios where self-interference is included, but with a potential performance degradation.}\cite{xiao2017full}. We employ uniform planar arrays (UPAs) with codebook-based beamforming as panel antenna arrays, each of which covers a $180^\circ$ area that is divided into $N_s$ sectors indicating possible beam directions. Each array panel is equipped with a single radio frequency (RF) chain thus able to create and process one baseband data stream, scheduling different single-beam at different panels. Therefore, an IAB-donor or IAB-node can process a total of $N_p$ baseband streams at a time. Fig. \ref{fig:principle}(b) provides an example of the top view of a node.
Based on the above designs, the IAB networks are deployed in the following.

In the \textit{backhaul}, two endpoints of a backhaul link point each other using the reciprocal sector and panel whose normal direction is the closest to the one of the LOS segment, as shown with the dashed lines in Fig.~\ref{fig:principle}(a) that depicts an example of an IAB wireless backhaul.
Nodes in the backhaul are equipped with buffers of unlimited size. As this work focuses only on the IAB networks, to eliminate the effect of traffic dynamics in the core network, we assume the IAB-donor's buffer to always store sufficient data\footnote{In this work, data flows are not differentiated and prioritized over different users, but regarded as equivalent for all users that are interested in the same data.}
to be delivered.
Each IAB-node holds a buffer storing the bits received via backhaul links from its parent. The buffers could be the bottleneck of multi-hop transmissions. \begin{changed}Indeed, if a buffer is empty, the activated links will transmit nothing and thus it causes a downlink starvation problem.\end{changed}
Therefore, the flow routing and link scheduling scheme is expected to timely refill the buffers to avoid any impact of the little amount of cached bits on downstream transmissions.

In the \textit{access}, UEs connecting to such a backhaul are associated to sectors based on their positions. A UE can belong to two sectors if it is located on a sector boundary. And UEs are expected to work in a dual-connectivity mode \cite{polese2017improved}, i.e., equipped with both legacy (3GPP FR1) and mmWave (3GPP FR2) interfaces. It is arguable that control-plane information can be exchanged through legacy FR1 interfaces to provide better coverage and signal propagation conditions.
Then, the control-plane FR1 interface can be used to send UE context information to enable better network access selection and configuration.
Therefore, we assume that each IAB-node is informed about associated UEs in real time and their channel status. Channel status information is typically available at each BS via Reference Signals and used for beamforming, rate adaptation, and other 5G procedures. BSs can also estimate the number and the position of connected UEs by activating network-side ranging techniques \cite{bernazzoli20235g}.
Instead, high-throughput user-plane channels can be established through mmWave (FR2) links.

The time domain $\mathcal{T}$ is divided into frames, each of which consists of $T$ slots of equal length $\delta$. 
The system, following a space-division multiple access (SDMA) scheme based on beamforming, takes advantage of the high directivity of mmWave antennas and can allow multiple concurrent transmissions, both backhaul and access links, to be carried out in each slot, by sharing the radio resources and carrying out transmissions through beams at different panels.
The simultaneous activation of several links requires the network to satisfy physical requirements, such as channel conditions (e.g., interference levels, antenna patterns), duplexing modes (i.e., FD, HD), RF chain limitations, UE hardware limitations, etc.
Signal-to-interference-plus-noise ratio (SINR) model is adopted to establish a successful link: a connection is created only if the SINR at the receiver is larger than the threshold required by the selected modulation and coding scheme (MCS). The interference that one link receives is the sum of the power it receives from all the non-intended transmitters simultaneously activated. Rate adaptation is considered as well, i.e., transmission rates depend on selected MCSs, which in turn depend on achievable SINR values.

\subsection{Network Operations}
\label{network_operations}

The transmissions follow and satisfy the rules and constraints below. A parent IAB-node -- more specifically, its RL agent(s) -- will have to choose between (1) sending data bits to a child IAB-node via a \textit{backhaul} link to refill its buffer, and (2) directly transmitting to a UE via an \textit{access} link to myopically improve its throughput. For a \textit{backhaul} transmission, if an IAB-node working in HD mode is selected as a receiver of its parent node in a specific slot, it cannot transmit in the same slot. For an \textit{access} transmission, a UE can be selected as a receiver if a beam points to its located sector. If more than one UE is present in the sector, one of them is randomly selected as the intended receiver.
Due to hardware limitations, a UE can receive from at most one IAB-node / IAB-donor in a slot, therefore, a collision occurs if a UE is selected as a receiver from more than one panel (i.e., more than one IAB-node) in the same time slot. Whenever it occurs, no bit can be delivered to the UE.
For both \textit{backhaul} and \textit{access} transmissions, if the amount of data bits cached in a buffer, rather than the link capacity, is the limiting factor, the outgoing links simultaneously activated equally share buffered bits to pursue the fairness.

The mmWave access network scenario above described is characterized by UEs moving with arbitrary directions and speeds, which may undergo link blockages caused by random mobile obstacles. This makes access links short-lived and unstable. In order to address such dynamics, in the next sections, we propose an adaptive MARL-based flow routing and link scheduling approach.

\setlength{\textfloatsep}{6pt}
\begin{table}[!h]
\scriptsize
    \caption{\small Summary of notations in Sections \ref{system_model} and \ref{method}.}
    \label{notation_1}
    \centering
    \begin{tabular}{@{\hspace{1pt}} l @{\hspace{1pt}} l}
        \toprule
        Notations & Definitions \\
        \midrule
        $\mathcal{G},\mathcal{V},\mathcal{E}$ & Network graph, node set, and edge set\\
        $\mathcal{R},\mathcal{U}$ & The set of IAB-nodes (including IAB-donor), and set of UEs\\
        $\mathcal{T},T,\delta$ & Time domain, the number of slots in a frame, and slot length\\
        $PL_{dB},G_{dB}$ & Path loss, and antenna gain\\
        $\phi_{-3dB}$ & Elevation HPBW\\
        $\beta_{-3dB}$ & Azimuth HPBW\\
        $N$ & The total number of RL agents (antenna array panels)\\
        $N_p,N_s$ & The No. of panels per IAB-node, and No. of sectors per panel\\
        $\mathcal{I}_p,\mathcal{I}_s$ & The sets of indicies for panels and sectors per panel\\
        $\mathcal{O},\mathcal{A},\pi$ & Observation space, action space and policy\\
        $o_t,a_t,r_t$ & Observation, action and reward at step $t$\\
        $o^{(i)}_t,r^{(i)}_t$ & The agent (panel) $i$'s observation and reward at step $t$\\
        $I^{pres}_{i,s}$ & Indicator of whether there are UEs in sector $s$ of agent (panel) $i$\\
        $A^{block}_{i,s}$ & Indicator of accumulated attenuation in sector $s$ of agent (panel) $i$\\
        $\mathcal{R}_i$ & The set of child IAB-nodes of agent (panel) $i$\\
        $L_n$ & The number of bits cached on the IAB-node $n$\\
        $B^N_n$ & The number of bits transmitted by the IAB-node $n$\\
        $B^P_i$ & The number of bits transmitted by the agent (panel) $i$\\
        $h(i)$ & The number of hops to reach agent (panel) $i$ from the IAB-donor\\
        $c_{min}$ & The minimum capacity available in the whole network\\
        $\rho_{BH}$ & The weight to counterbalance the large backhaul link capacity\\
        $\mathcal{I}_p^{EB}$ & The set of
panels with empty buffers\\
        $\mathcal{I}_p^{RX}$ & The set of panels whose located IAB-nodes are receiving data\\
        $\mathcal{I}_p^{ER}$ & The union of $\mathcal{I}_p^{EB}$ and $\mathcal{I}_p^{RX}$\\
        $\zeta$ & The penalty term in the reward function \\
        \bottomrule
    \end{tabular}
\end{table}

\vspace{-10pt}
\section{Adaptive Flow Routing and Link Scheduling}
\label{method}
A first approach to apply RL\footnote{A brief introduction to RL can be found in Appendix Sec. A.} to mmWave IAB networks is to consider a central network controller that acts as a single RL agent. However, this requires the agent to know the global state of the whole network and manage all the transmissions, resulting in a combinatorially large number of different states and possible actions, which increases exponentially with the size of the network.
It becomes even worse when mobile UEs and obstacles are introduced.
This would strongly limit the scalability and the flexibility of the resource allocation approach.
These reasons motivated us to resort to the MARL technique that allows to split the overall complexity into several smaller problems managed by cooperative agents.

We consider multiple RL agents and assign each to an IAB-node / IAB-donor Tx array panel, such that each agent controls the beamforming direction of the associated Tx panel in each time slot.
Each Tx panel (RL agent) cooperates with the other Tx panels to learn the environment dynamics and understand the impact of the other Tx panels' policies.
Their collective goal is to maximize the throughput (namely, the number of bits per frame) delivered to UEs. This requires a proper management of the backhaul and access link transmissions during a frame.
Note that one time slot of the frame corresponds to one step in the RL interactions, and one episode in the RL interactions is equivalent to one frame. Each agent executes an action in each slot, according to the policy $\pi$ available at the beginning of the slot.
How to achieve high UE throughput without significantly reducing fairness depends on how Tx antenna panels (agents) point their beams, how IAB-node buffers are refilled, and how data bits flow through the network, crossing IAB-nodes.
These aspects will be managed by the MARL agents trained based on the observation, action and reward components designed below.

\vspace{3pt}
Considering $|\mathcal{R}|$ IAB-nodes (including the IAB-donor), each equipped with $N_p$ Tx antenna panels, there are a total number of $N=|\mathcal{R}|\cdot N_p$ agents in the scenario, indexed by $\mathcal{I}_p = \{ 1, \dots, N\}$. Each agent faces $N_s$ sectors indexed by $\mathcal{I}_s = \{1, \dots, N_s\}$. The observation space $\mathcal{O}$, action space $\mathcal{A}$, and reward function of the RL agents are defined as follows, considering IAB-nodes operating in FD and HD modes.

\vspace{-7pt}
\subsection{Observation Space}
\vspace{-3pt}

We design different observation vectors for RL agents in FD and HD IAB networks, respectively.
They contain a duplex-mode-specific element and several other common elements.

\textbf{FD Mode:}
For each agent $i$, the observation includes the information of:

\noindent\emph{a) UE presence} - $I^{pres}_{i, s}$, which takes value $1$ for sector $s$ of agent $i$ if there are UEs located under the coverage of sector $s$; $0$, otherwise. This information can be easily estimated by using signaling and context information (e.g., position, received power, etc.) sent by UEs. An example of this binary information is shown in Fig. \ref{fig:principle}(c).

\noindent\emph{b) Sector attenuation} - $A^{block}_{i, s}$, which is the average additional attenuation over the path loss model experienced by the transmissions carried out in sector $s$ of agent $i$ in the previous frame. We can assume that this attenuation is caused by an interposing obstacle.
Blockers moving at realistic speeds can lead to sector obstructions
that last hundreds or even thousands of slots and suddenly disappear in few slots with the departure of the blockers \cite{maccartney2017rapid}. A similar behavior is repeated for sectors in the opposite transition (from no obstruction to obstruction). Therefore, the obstruction status in the previous slots can provide a reliable reference for the current slot. Indeed, the number of slots in which a transition happens is very small and the instants they occur are very hardly predictable in any case.
	
\noindent\emph{c) Child-node buffer level} - $L_n$, which indicates the amount of data bits cached in the buffer of the child IAB-node $n$, reachable through Tx panel $i$.
Such information is essential for parent IAB-donor/IAB-node agents to plan their buffer-refilling strategies.

The above information is organized in concatenated sub-vectors to form an observation vector for agent $i \in \mathcal{I}_p$:
\begin{equation}
	o^{(i)}_t = {[[I^{pres}_{i, s}]_{s \in \mathcal{I}_s}, [A^{block}_{i, s}]_{s \in \mathcal{I}_s}, [L_n]_{n \in \mathcal{R}_i}]},
	\label{observation}
\end{equation}
where $\mathcal{R}_i$ is the set of child IAB-nodes reachable via agent $i$.

\textbf{HD Mode:}
All the above three sub-vectors also appear in the observation vector for the HD mode. However, differently from FD mode, operating in HD mode introduces restrictions between parent and child nodes: if a parent node transmits to a child node, the receiving child node cannot simultaneously transmit. This may hinder the data delivery in the tree topology where data bits require multiple backhaul hops to reach UEs. This poses a big challenge to the cooperation of RL agents, which have to coordinate several concurrent and dynamic factors (i.e., interference reduction, buffer refill and collision avoidance).

Therefore, we add a fourth sub-vector, $[B^N_{n}]_{n \in \mathcal{R}_i}$, to the observation vector, which includes the amount of data bits transmitted downstream by every child IAB-node $n$ reachable through the agent $i$ ($n \in \mathcal{R}_i$). This information, together with $L_n$, allows the parent agent to balance between the buffer level and the transmission opportunity of each child node in order to both avoid empty buffers and provide a good downstream throughput. As a result, the observation vector of the agent $i\in\mathcal{I}_p$ working in HD mode is:
\begin{equation}
	o^{(i)}_t = {[[I^{pres}_{i, s}]_{s \in \mathcal{I}_s}, [A^{block}_{i, s}]_{s \in \mathcal{I}_s}, [L_n]_{n \in \mathcal{R}_i}, [B^N_n]_{n \in \mathcal{R}_i}]}
	\label{observation}
\end{equation}

\vspace{-15pt}
\subsection{Action Space}

Each agent $i \in \mathcal{I}_p$, working in either HD or FD mode, can choose to (1) activate one of its $N_s$ sectors to transmit to a covered UE ($ACC$ action), to (2) transmit to one of the reachable child IAB-nodes in the set $\mathcal{R}_i$ ($BH$ action), or to (3) stay silent ($SIL$ action).

The sector-based access transmissions in (1) allow to reduce the impact of the varying UE locations on the action policies, making them more stable and robust against mobility. In addition, this permits the same action space to remain widely applicable even if the number of UEs in the service area may not be constant. Indeed, once a sector is selected, only one UE, if any, will be randomly selected to be served.

In each slot, every agent tries to establish a link according to the selected action. Concurrent links interfere with each other, hence their delivered data amount depends on experienced SINR values. Whether or not activated links can truly deliver that number of bits finally depends on whether IAB nodes have enough bits buffered and whether blockages are caused by obstacles.
The target of maximizing the UE throughput can be achieved by properly tuning the rewards for actions, as indicated in the following reward functions.

\vspace{-12pt}
\subsection{Reward Function}

We design two reward functions for the FD and HD cases, respectively. Similarly to the definition of the observation space, the reward function for the HD case shares some common elements with the one for the FD case and has an additional element to prevent IAB-node buffer starvation.

\textbf{FD Mode:}
Considering the three types of actions aforementioned, we discuss the definition of the reward function.

If agent $i$ chooses $ACC$ action and succeeds in serving a UE, it gets a positive reward equal to the number of bits $B^P_i$ sent by the Tx panel $i$, multiplied by the factor $(h(i)+1)$ and normalized by $c_{min}$, the minimum capacity (corresponding to the minimum MCS) available in the whole network. The term $h(i)$ is the number of hops separating the IAB-node, where the agent $i$ is located, from the IAB-donor. Therefore, the factor $h(i)+1$ is applied to give higher rewards to transmissions serving UEs farther away from the IAB-donor. This facilitates the use of the backhaul resources rather than relying on the myopic traffic delivery to nearby UEs, especially those directly connected to the IAB-donor. Since backhaul links typically have higher MCS values than access links, favoring the use of IAB-nodes allows to increase the capacity of the network. In addition, IAB-nodes are essential to increasing the number of covered users.

If agent $i$ executes $BH$ action and successfully transmits data via a backhaul link, similarly to the $ACC$ action, the reward is proportional to the amount of transferred data bits $B^P_i$ normalized by $c_{min}$ and multiplied by $(h(i)+1)$. In addition, this fraction is further multiplied by a weight $\rho_{BH}\in(0,1)$ that is used to counterbalance the fact that the link capacity, and thus the number of transferred bits, in backhaul is usually remarkably larger than that in access. Without $\rho_{BH}$, agents would largely prefer to activate backhaul links, accumulating data bits in IAB-nodes' buffers.

If the agent $i$ chooses $ACC$ or $BH$ action and the transmission fails due to either an empty IAB-node buffer or a collision at a UE, the agent $i$ will get a penalty $-\zeta$ for its neglecting the buffer length or not cooperating well with the other partner agents.

When agent $i$ plays $SIL$ action, there can be two types of outcomes. 1) If coincidentally, the buffer of the IAB-node, where agent $i$ is installed, is empty ($i \in \mathcal{I}_p^{EB}$), which is an external limitation not directly ascribable to the agent's policy, the agent $i$ gets a reward 0 to prevent the training process from being biased by the empty buffer. 2) If there are data bits in the buffer, the agent $i$ will get a penalty $-\zeta$ (empirically, $\zeta=1$) in order to incentivize policies that increase the throughput. Therefore, the reward of agent $i$ working in the FD mode is:
	\begin{equation}
		r^{(i)}_t = \left\{ \begin{array}{lll}
			 \frac{(h(i)+1) \cdot B^P_i}{c_{min}}, & \text{if } B^P_i > 0, ACC \mbox{ act.},\\
			\rho_{BH} \cdot \frac{(h(i)+1)\cdot B^P_i}{c_{min}}, & \text{if } B^P_i> 0, BH \mbox{ act.},\\
			0, & \text{if } i \in \mathcal{I}^{EB}_p, SIL \mbox{ act.}\\
			-\zeta, & \text{otherwise}.
		\end{array}\right.
		\label{reward_iab_donor}
	\end{equation}

\textbf{HD Mode:} The reward function for the HD IAB networks is the same as that for the FD IAB networks, except for the following two aspects.

The reward for the agent $i$, thanks to its successful backhaul transmission to an IAB-node $n$, is additionally scaled by the IAB-node $n$'s buffer length $L_n$ after the transmission is completed. This additional scaling factor gives smaller rewards to those transmissions whose receiving nodes have accumulated a large number of bits in their buffers. This offers two benefits. First, child nodes will not experiment situations where they constantly receive from their parent nodes, which, given HD constraints, would prevent them from transmitting downstream to other nodes. Second, the variance of buffer lengths across different IAB-nodes can be reduced. This can reduce the UE throughput variance, thus contributing to a better UE throughput fairness.

If the agent $i$ chooses the $SIL$ action, the reward is set to $0$ not only when its IAB-node's buffer is empty ($i \in \mathcal{I}_p^{EB}$), but also when its IAB-node is receiving from a parent node in the same slot ( $i \in \mathcal{I}_p^{RX}$). To simplify the notation, we define $\mathcal{I}_p^{ER} = \mathcal{I}_p^{EB} \cup \mathcal{I}_p^{RX}$. Therefore, the reward of agent $i$ working in HD mode is written as follows.
\begin{equation}
    r^{(i)}_t = \left\{ \begin{array}{lll}
        \frac{(h(i)+1) \cdot B^P_i}{c_{min}}, & \text{if } B^P_i > 0, ACC \mbox{ act.},\\
        \rho_{BH} \cdot \frac{(h(i)+1)\cdot B^P_i}{c_{min}\cdot L_n}, & \text{if } B^P_i> 0, BH \mbox{ act.},\\
        0, & \text{if } i \in \mathcal{I}^{ER}_p, SIL \mbox{ act.},\\
        -\zeta, & \text{otherwise}.
    \end{array}\right.
    \label{reward_iab_donor}
\end{equation}

\begin{table}[!t]

\scriptsize
    \caption{\small Summary of notations in Sections \ref{learning_approach} and \ref{implementation_issue}.}
    \label{notation_2}
    \centering
    \begin{tabular}{@{\hspace{1pt}} l @{\hspace{1pt}} l}
        \toprule
        Notations & Definitions \\
        \midrule
        $\mathbf{o}_t,\mathbf{a}_t,\mathbf{r}_t$ & Vectors of $N$ agents' observations, actions and rewards\\
        $o^{(i)}_t,a^{(i)}_t,r^{(i)}_t$ & The observation, action and reward of agent $i$ at step $t$\\
        $\pi_{\theta^{(i)}},\theta^{(i)}$ & Agent $i$'s policy function and its parameter\\
        $\pi_{\boldsymbol{\uptheta}},\boldsymbol{\uptheta}$ & The collection of $N$ agents' policies and their parameters\\
        $\hat{a}^{(i)}_t$ & Agent $i$'s action sampled from policy $\pi_{\theta^{(i)}}$ at step $t$\\
        $\mathbf{\hat{a}}_t$ & The vector of $N$ agents' actions sampled from $\pi_{\boldsymbol{\uptheta}}$ at step $t$\\
        $\mathbf{\hat{a}}^{(\backslash i)}_t$ & The vector of $N$ agents' actions (except $i$'s) at step $t$\\
        $\pi_{\bar{\theta}^{(i)}},\pi_{\boldsymbol{\bar{\uptheta}}}$ & Agent $i$'s target policy and the collection of $N$ agents'\\
        $\bar{a}^{(i)}_t$ & Agent $i$'s action sampled from the target policy $\pi_{\bar{\theta}^{(i)}}$ at step $t$\\
        $\mathbf{\bar{a}}_t$ & The vector of $N$ agents' actions sampled from target policies\\
        $Q^{\psi,(i)}$ & The Q-value function corresponding to agent $i$\\
        $f^{(i),e^{(i)}}$ & The dedicated parts in Q-value DNN model for agent $i$\\
        $g,\omega^{(j)}$ & The shared parts in Q-value DNN model among agents\\
        $x^{(\setminus i)}$ & The other agents' contribution to agent $i$'s Q-value\\
        $Q^{\bar{\psi},(i)}$ & The target Q-value function corresponding to agent $i$\\
        $\tau$ & Trade-off weight in the Q-value loss function\\
        $\gamma$ & Discount factor for rewards\\
        $T_{upd}$ & The number of data entries sent by each agent at each time\\
        $Tup^{(i)}$ & The batch of tuples sent by agent $i$ to the central entity\\
        $tup^{(i)}$ & A tuple in the batch $Tup^{(i)}$ sent by agent $i$\\
        $h,h_{size}$ & The mini-batch sampled from $H$ and its size\\
        $A$ & The number of bits used to record the attenuation for a sector\\
        $L$ & The number of bits used to indicate the buffer level\\
        $B$ & The number of bits used to record the amount of data delivered\\
        $N^{ch}_i$ & The number of child IAB-nodes connected to agent $i$\\
        $T^A_l,T^C_l$ & The transmission latency of the agents and central entity\\
        $K$ & The number of consecutive DNN updates carried out each time\\
        $\kappa$ & The moving average weight in updating $\bar{\psi}$ and $\bar{\theta}$\\
        $T_{wup}$ & The number of steps in the warm-up period\\
        \bottomrule
    \end{tabular}
\end{table}

\begin{figure*}[ht]
	\centering
	\includegraphics[width=0.8\textwidth]{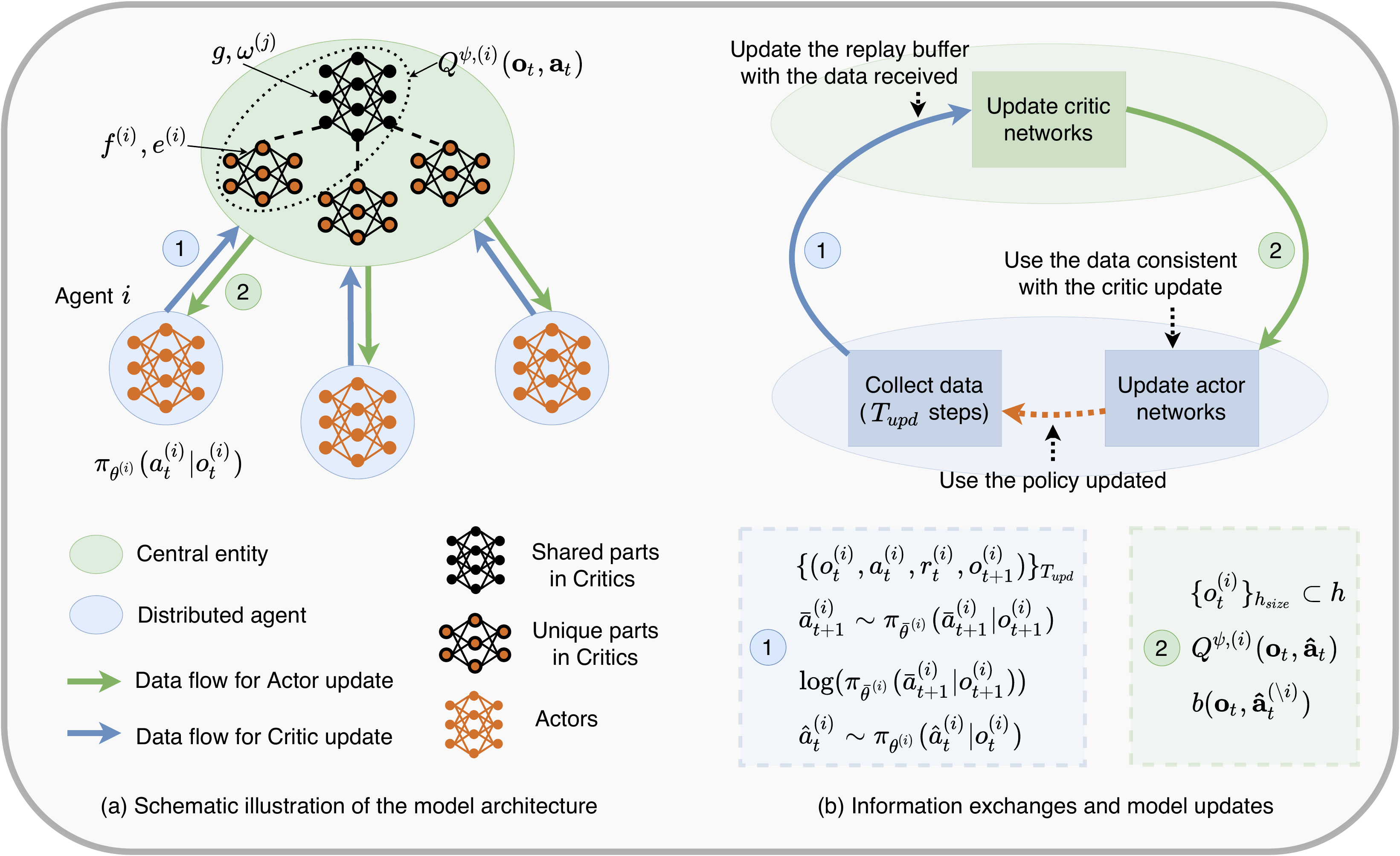}
	\caption{\small Basic principles of the MARL framework: (a) an overview of the MARL system architecture, (b) key components of information exchanges and model updates.}
	\label{fig:com_coop_combined}
    \vspace{-5mm}
\end{figure*}

Cooperation is fundamental to the effective learning of the agents formulated above. Simply applying independent SARL algorithms to train individual agents interprets the other agents' decisions as part of the environment, which would be, in turn, non-stationary as the other agents' policies constantly change as well during the learning process. Therefore, the MARL algorithm is utilized for training purposes.

\section{Reinforcement Learning Approach}
\label{learning_approach}

To effectively train the RL agents whose components have been designed in Sec. \ref{method}, we propose a learning framework in Sec.~\ref{implementation_issue}, based on a Multi-Actor-Attention-Critic (MAAC) approach \cite{iqbal2019actor}.
The RL agents are trained to pursue a collective goodness, by leveraging two remarkable advantages:
(1) its critic part allows each agent to automatically consider observations and actions only from relevant agents (based on the idea of \textit{attention}), thus filtering out information not correlated to a performance improvement; (2) it trains decentralized policies with centrally-computed critics, which allows agents to possess individual policies and independently apply them, once training is completed.

\textbf{Attention-Based Critics} Denoting observations, actions, rewards at step $t$ as respectively $\mathbf{o}_t=(o^{(1)}_{t},\dots,o^{(N)}_{t})$, $\mathbf{a}_t=(a^{(1)}_{t},\dots,a^{(N)}_{t})$, $\mathbf{r}_t=(r^{(1)}_{t},\dots,r^{(N)}_{t})$ and the policy parameters of all the $N$ agents as $\boldsymbol{\uptheta}=(\theta^{(1)},\dots,\theta^{(N)})$,
the rationale is to train each agent $i$'s individual policy $\pi_{\theta^{(i)}}(a^{(i)}_t|o^{(i)}_t)$ by means of its action-value $Q^{\psi,(i)}(\mathbf{o}_t,\mathbf{a}_t)$. The value of $Q^{\psi,(i)}(\mathbf{o}_t,\mathbf{a}_t)$ is centrally computed by using information from other relevant agents, according to an attention mechanism. Therefore, agent $i$'s $Q^{\psi,(i)}(\mathbf{o}_t,\mathbf{a}_t)$ depends not only on its own observation $o^{(i)}_t$ and action $a^{(i)}_t$, but also on those of the other agents, combined as follows:
\begin{align}
	&Q^{\psi,(i)}(\mathbf{o}_t,\mathbf{a}_t) = f^{(i)}(e^{(i)}(o^{(i)}_t, a^{(i)}_t), x^{(\setminus i)}), \label{q_function}\\
	&x^{(\setminus i)} = \sum\nolimits_{j\neq i}\omega^{(j)} \cdot g(e^{(j)}(o^{(j)}_t,a^{(j)}_t)), \label{others}
\end{align}
where $f^{(i)}$ consists of a fully connected layer with leaky ReLU and a linear layer, while $e^{(i)}$ is an embedding function implemented as a fully connected layer with leaky ReLU. The contribution $x^{(\setminus i)}$ to agent $i$ from the other agents is a weighted sum of functions of other agents' embedding functions. In particular, $g$ is a fully connected layer with leaky ReLU. Weight $\omega^{(j)}$ is the \textit{attention weight} associated to the information provided by agent $j$, which is optimized during the training phase, together with the other weights of the neural network. The value of $\omega^{(j)}$ is determined by the similarity between $e^{(i)}$ and $e^{(j)}$, which is computed using a matching approach between ``query" based on $e^{(i)}$ and ``key" based on $e^{(j)}$.
In Eqs. \eqref{q_function} and \eqref{others}, $f^{(i)}$ and $e^{(i)}$ are \emph{dedicated parts} to each agent, while $g$ and $\omega^{(j)}$ are \textit{shared parts} among all the agents. They can be visualized in Fig.~\ref{fig:com_coop_combined}(a). The centralized action-value function $Q^{\psi,(i)}(\mathbf{o}_t,\mathbf{a}_t)$ of agent $i$, encircled with a black dot line, includes the black shared critic NN and an orange black dedicated critic NN.

\textbf{Model Updates} As mentioned above, functions and parameters in Eqs. \eqref{q_function} and \eqref{others} are implemented as DNNs, whose weight vector $\psi$ is updated considering a replay buffer $H$ that stores history trajectories in the form of $(\mathbf{o}_t, \mathbf{a}_t, \mathbf{r}_t, \mathbf{o}_{t+1})$ entries, which summarize the interactions occurred in the previous steps. In particular, based on a set of entries randomly sampled from $H$, we update $\psi$ to minimize the following joint regression loss function:
\begin{equation}
	\mathcal{L}_Q(\psi) =  \sum^{N}_{i=1}\mathbb{E}_{(\mathbf{o}_t,\mathbf{a}_t,\mathbf{r}_t,\mathbf{o}_{t+1})\sim H}[(Q^{\psi,(i)}(\mathbf{o}_t,\mathbf{a}_t)-y^{(i)}_t)^2], \label{critic_loss}
\end{equation}
\begin{align}
    y^{(i)}_t = & r^{(i)}_t + \gamma\mathbb{E}_{\mathbf{\bar{a}}_{t+1}\sim\pi_{\boldsymbol{\bar{\uptheta}}}(\mathbf{o}_{t+1})}[Q^{\bar{\psi},(i)}(\mathbf{o}_{t+1}, \mathbf{\bar{a}}_{t+1}) - \notag\\
    & \tau\log(\pi_{\bar{\theta}^{(i)}}(\bar{a}_{t+1}^{(i)}|o_{t+1}^{(i)}))],\label{y_i}
\end{align}
where $Q^{\bar{\psi}}$ and $\pi_{\boldsymbol{\bar{\uptheta}}}$, respectively called target action-value functions and target policies, are moving averages of the past action-value and policy functions (used to stabilize the training), while $\bar{a}_{t+1}^{(i)}$ is the next action that agent $i$ would select by applying the target policy $\pi_{\bar{\theta}^{(i)}}$ to the next observation $o_{t+1}^{(i)}$. 
The logarithmic term in Eq. \eqref{y_i} is the policy entropy. It encourages action-space exploration by promoting random selections, thus reducing the probability to converge to deterministic policies with poor local optima.
Parameter $\tau$ is the trade-off weight used to balance the importance of reward maximization over random exploration.
Finally, note that the evaluation of this loss function requires a joint optimization of $\psi$ across all the individual action-value functions $Q^{\psi,(i)}(\mathbf{o}_t,\mathbf{a}_t)$ ($i\in\mathcal{I}_p$), therefore it is performed at the central entity.

Once $Q^{\psi,(i)}(\mathbf{o}_t,\mathbf{a}_t)$ is updated using Eq. \eqref{critic_loss}, the individual policy $\pi_{\theta^{(i)}}(a^{(i)}_t | o^{(i)}_t)$ of each agent $i$ (shown as an orange actor NN in Fig.~\ref{fig:com_coop_combined}(a)) can be updated as well, according to a gradient ascent approach over DNN weights $\theta_i$:

\begin{equation}
    \nabla_{\theta^{(i)}}J(\pi_{\boldsymbol{\uptheta}}) = \mathbb{E}_{\mathbf{o}_t\sim H, \mathbf{\hat{a}}_t\sim \pi_{\boldsymbol{\uptheta}}} \left[\vphantom{Q^{\psi,(i)}}
	\nabla_{\theta^{(i)}} \log( \pi_{\theta^{(i)}}(\hat{a}^{(i)}_t|o^{(i)}_t ))\cdot \right. \nonumber
 \end{equation}
 \begin{equation}
    \left.\left( -\tau\log( \pi_{\theta^{(i)}}(\hat{a}^{(i)}_t|o^{(i)}_t) )+
	Q^{\psi,(i)}(\mathbf{o}_t,\mathbf{\hat{a}}_t) - b(\mathbf{o}_t,\mathbf{\hat{a}}^{(\backslash i)}_t) \right) \right].
	\label{policy_gradient}
\end{equation}
Each policy network with $\theta^{(i)}$ is composed of three linear layers and a final leaky ReLU.
The baseline $b\left(\mathbf{o}_t,\mathbf{\hat{a}}^{(\backslash i)}_t\right)$ allows to reduce the variance of the gradient and is computed by averaging $Q^{\psi,(i)}(\mathbf{o}_t,\mathbf{\hat{a}}_t)$ (i.e., $Q^{\psi,(i)}(\mathbf{o}_t,(\hat{a}^{(i)}_t,\mathbf{\hat{a}}^{(\backslash i)}_t))$) over all possible actions of agent $i$ (i.e., $\hat{a}^{(i)}_t$) according to the policy distribution $\pi_{\theta^{(i)}}$, keeping fixed the actions of the other agents (i.e., $\mathbf{\hat{a}}^{(\backslash i)}_t$). Note that the update of $\theta^{(i)}$, although based on a per-agent action-value $Q^{\psi,(i)}(\mathbf{o}_t,\mathbf{\hat{a}}_t)$, requires the knowledge of the other agents' observations sampled from $H$ and the other agents' action probabilities expressed by $\pi_{\boldsymbol{\uptheta}}$.

\vspace{-10pt}
\section{Learning Framework for IAB Networks}
\label{implementation_issue}
\label{com_coo_scheme}

This section delineates the model architecture for mmWave IAB networks and scrutinizes the significant challenges encountered during the training process. To effectively tackle these challenges, we introduce a training cycle synchronization scheme, which facilitates the scheduling of multi-agent training procedures, considering practical aspects highlighted by challenges.
And we present in detail the training process, including message exchanges, for the central entity and agents.

\vspace{-10pt}
\subsection{Model Deployment in IAB Networks}

\textbf{Model Architecture}
The centralized critics are computed at a central entity located at the IAB-donor, while local policies are distributed at agents associated to Tx antenna panels at both IAB-donor and IAB-nodes. The centralized critics (i.e., the DNN in the green circle in Fig.~\ref{fig:com_coop_combined}(a)) act as a bridge among local policies and implicitly capture the agents' cooperation during training phase. The training of such a semi-distributed architecture relies on message exchanges between the IAB-donor and IAB-nodes, which can be carried out through direct control-plane links working at FR1 frequencies. As we will show later, only a limited amount of information has to be exchanged to achieve good results.
And since agents no longer need central critics once training is concluded, leaving the operation phase with fully distributed and independently policies, message exchanges are required only during the training phase. The key components of model updates and message exchanges are shown in Fig.~\ref{fig:com_coop_combined}(b).

\textbf{Training Challenges}
Several issues will arise when mmWave IAB networks perform training procedures practically, because message exchanges between the IAB-donor and IAB-nodes are required during training.
The message exchanges are unavoidably affected by non-negligible latencies, which can deteriorate the training process in the following two ways. First, the latencies can slow down the learning process. Second, different IAB-nodes can experience distinct latencies due to their different distances from the IAB-donor and thus incur coordination issues and other inconvenience for the central entity and distributed agents. A typical case is that the messages from different agents can arrive at the central entity at different moments. This forces the central entity to wait for messages from remote agents to guarantee consistent experience trajectories to be stored in the replay buffer. This will greatly slow down the training process. In turn, agents can receive messages from the central entity at different moments, causing coordination issues among agents. We address the above issues in the following sections.

\vspace{-10pt}
\subsection{Training Cycle Synchronization}
\begin{figure}[!t]
    \centering
    \includegraphics[width=0.49\textwidth]{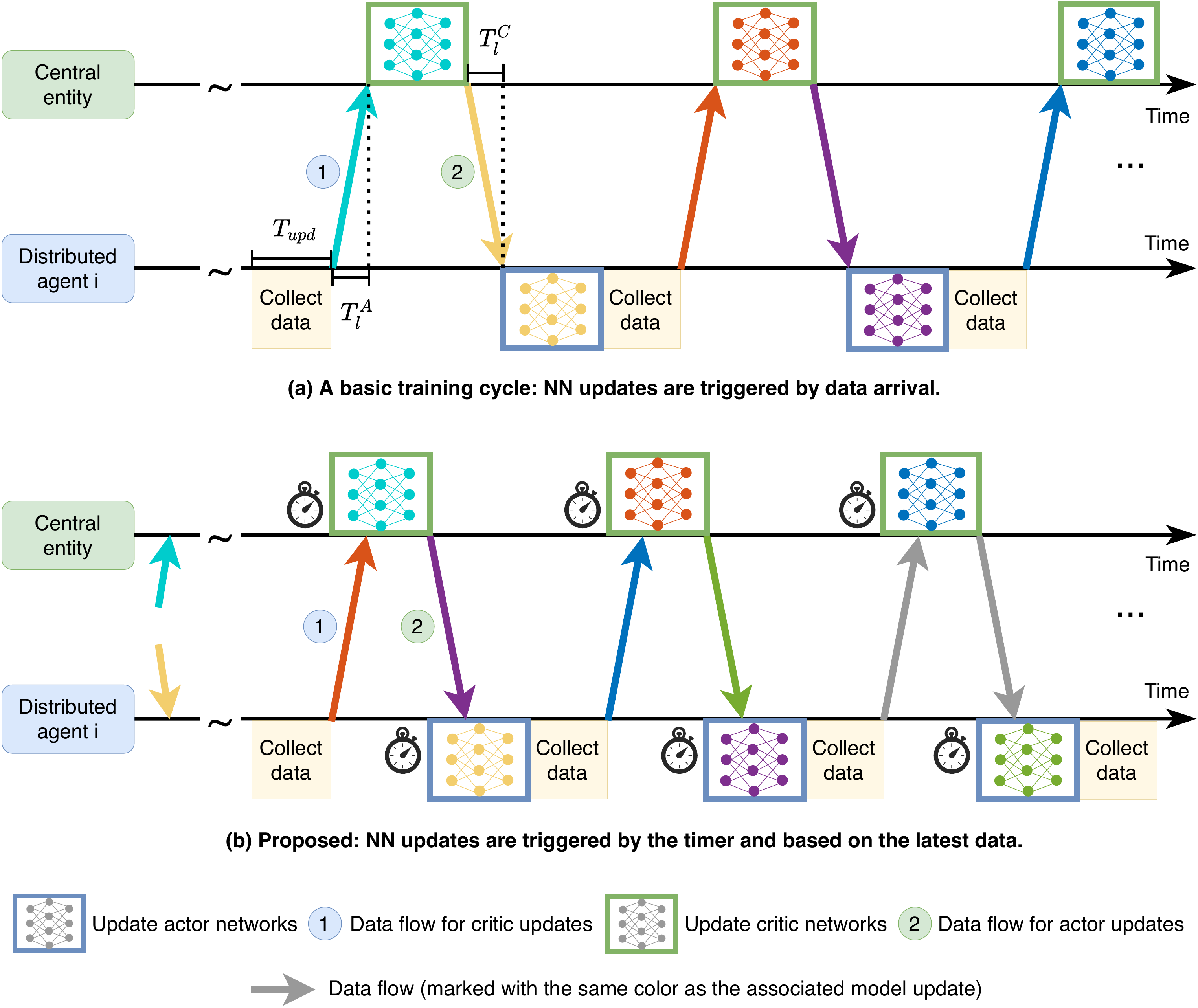}
    \caption{\small An illustrative example of system coordination solutions.}
    \label{fig:late_updates}
    \vspace{-3pt}
\end{figure}

A basic training cycle consists of the following steps, as shown in Fig.~\ref{fig:late_updates}(a). The agents collect experience data by performing network operations under the existing policies, and send these data together with some local policy information to the central entity. The central entity puts the received data in the replay buffer, samples the training data, and updates the centralized critics. Subsequently, the data required in policy updates is sent to the agents, which update their policies and generate new experience data with the latest updated policies. The above steps repeat to form the basic training cycle.
However, such a training cycle is characterized by the information exchange latency (as indicated by $T_l^A$ and $T_l^C$ in Fig.~\ref{fig:late_updates}(a)), which can significantly decrease the training efficiency and deteriorate the coordination of the whole system.

Therefore, we present a synchronized training cycle, as shown in Fig.~\ref{fig:late_updates}(b), which can be summarized in following two aspects.
First, the updates at both the central entity and distributed agents are performed periodically, based on pacing timers.
The timers are appropriately configured to take into account the central entity and agents and coordinate the model updates and message exchanges of all the entities at the same pace.
Second, central critic updates are performed with the data sampled from the most recent available content of the replay buffer, while distributed agents' policies are updated with the latest available data received from the central entity.
This idea is illustrated by the corresponding DNN updates and information transmissions with matching colors.

\vspace{-10pt}
\subsection{Overview of the Training Process}

The training process consists of a sequence of update instants, in each of which, multiple DNN updates are performed by sampling $K$ random mini-batches from the replay buffer and updating DNNs' weights accordingly. Moreover, an initial transient period is needed to reach a steady state (i.e., stationary buffer levels and link behaviors), therefore a warm-up period (e.g., the first $T_{wup}$ steps) is considered, during which, no update is performed.

Algorithm \ref{learning_alg_central} details the learning procedure for the central entity. 
(1) [Lines 3-6] Every time the central entity detects the arrival of the new experience information from agents, the tuples contained in the message are merged into the replay buffer. 
(2) [Lines 8-12] When the timer expires (and the warm-up period is concluded), $K$ random mini-batches are sampled and used to update the Q-value function. 
(3) [Lines 13-16] The updated Q-value function is used to compute new policy-update information, which is sent to each agent. 

Algorithm \ref{learning_alg_each_agent} presents the learning procedure for each agent $i$. 
(1) [Lines 3-5] Every time the agent $i$ receives new information from the central entity, it collects information for $K$ consecutive updates.
(2) [Lines 7-10] When the timer expires (and the warm-up period is concluded), $K$ consecutive policy function updates are performed.
(3) [Lines 11-12] Using the updated policy, the agent $i$ interacts with the environment to collect the experience trajectory data and computes the values of associated variables.  
(4) [Line 13] Agent $i$ sends experience tuples and other related information to the central entity.

The time between two updates (i.e., update-timer interval) directly impacts the convergence speed, but it is potentially arbitrary. Indeed, it can be set according to the length of an arbitrary episode, the time to process an update, or the replay-buffer sampling factor to generate a mini-batch. However, to strike a balance between model training efficiency and message exchange costs, a proper update interval is needed.

Once the convergence is reached (f.i., after a maximum number of steps or when minimal NN weight updates are performed), the training procedure stops and distributed agents continue the interaction with the environment based on local observations and fixed local policies. However, the training procedures can be re-activated at any point in time during the system operation, f.i., periodically or when a substantial performance decrease is detected. Indeed, new information collected at the agents can be accumulated and sent to the central entity at any time. Likewise, critic updates and policy updates can be performed at any time, when a sufficient number of new tuples have been inserted in the replay buffer.

\vspace{-10pt}
\subsection{Training Details for Central Entity}

The whole training phase is based on the update of the central action-value functions $Q^{\psi,(i)}(\mathbf{o}_t,\mathbf{a}_t)$ ($i\in\mathcal{I}_p$) in Algorithm \ref{learning_alg_central} [Line 11], which plays an essential role in enabling cooperation among distributed agents. 
This task requires the central entity to have a centralized replay buffer $H$ storing the past experiences of every agent.
	
During network operations, agent $i$ plays action $a^{(i)}_t$ at each step $t$, which is selected by using its own policy $\pi_{\theta^{(i)}}$ on the basis of its observation $o^{(i)}_t$. This leads to a reward $r^{(i)}_t$ and a new observation $o^{(i)}_{t+1}$. New tuples $(o^{(i)}_t, a^{(i)}_t, r^{(i)}_t, o^{(i)}_{t+1})$ are temporarily and locally accumulated at each agent $i$, which periodically sends a tuple batch $Tup^{(i)}=\{(o^{(i)}_t, a^{(i)}_t, r^{(i)}_t, o^{(i)}_{t+1})\}_{T_{upd}}$ to the central entity considering the last $T_{upd}$ steps. The central entity merges received per-agent tuples into agent-indexed vector tuples $\{(\mathbf{o}_t,\mathbf{a}_t,\mathbf{r}_t,\mathbf{o}_{t+1})\}_{T_{upd}}$ and inserts them into $H$.

In each of the $K$ iterations of a periodic update, the central entity samples a random mini-batch $h$ from $H$ consisting of a number $h_{size}$ of vector tuples (i.e., $h=\{(\mathbf{o}_t, \mathbf{a}_t, \mathbf{r}_t, \mathbf{o}_{t+1})\}_{h_{size}}$) and uses it in the minimization of the joint regression loss to update $Q^{\psi,(i)}(\mathbf{o}_t,\mathbf{a}_t)$ ($i\in\mathcal{I}_p$). Namely, for each vector tuple in $h$, the central entity updates $\psi$ according to the following equations:
\begin{align}
    & \mathcal{L}_Q(\psi) = \sum_{i \in \mathcal{I}_p}[(Q^{\psi,(i)}(\mathbf{o}_t,\mathbf{a}_t)-y^{(i)}_t)^2], \label{critic_loss_alg}\\
    & y^{(i)}_t = r^{(i)}_t + \gamma[Q^{\bar{\psi},(i)}(\mathbf{o}_{t+1}, \mathbf{\bar{a}}_{t+1}) - \tau\log(\pi_{\bar{\theta}^{(i)}}(\bar{a}_{t+1}^{(i)}|o_{t+1}^{(i)}))].\label{y_i_alg}
\end{align}
Note that, while mini-batch $h$ is generated at the central entity, $\bar{a}_{t+1}^{(i)}$ and the value of $\log(\pi_{\bar{\theta}^{(i)}}(\bar{a}_{t+1}^{(i)}|o_{t+1}^{(i)}))$ are received from remote agents together with the tuple in $h$.
After the update of $Q^{\psi,(i)}$ ($i\in\mathcal{I}_p$), target action-value functions $Q^{\bar{\psi},(i)}$ ($i\in\mathcal{I}_p$) are updated using a moving average with an update rate $\kappa$.

\begin{algorithm}[t]
    \small
    \caption{\small\textit{Learning Procedures for Central Entity}}
    \label{learning_alg_central}
    Parameters: $T_{train}$, $T_{upd}$, $T_{wup}$, $K$, $h_{size}$, $\kappa$.
    \begin{algorithmic}[1]
        \State Initialize $\psi$, $H$;
        \For{$t_{train}=1, \dots, T_{train}$}
            \vspace{6pt}
            \State \underline{\textbf{\textit{Data Arrival Detection:}}}
            \For{$i \in \mathcal{I}_p$} \Comment{Arrive at different time due to latency.}
                \State \textbf{\textit{Read}} the set of tuples $\{(o^{(i)}_t,a^{(i)}_t,r^{(i)}_t,o^{(i)}_{t+1})\}_{T_{upd}}$ and \Statex ~~~~~~~~~~~~~~corresponding values $\hat{a}^{(i)}_t$, $\bar{a}^{(i)}_{t+1}$ and 
                \Statex ~~~~~~~~~~~~~~$\log(\pi_{\bar{\theta}^{(i)}}(\bar{a}_{t+1}^{(i)}|o_{t+1}^{(i)}))$ from agent $i$;
            \EndFor
            \State $H \gets H \cup \{(\mathbf{o}_t, \mathbf{a}_t, \mathbf{r}_t, \mathbf{o}_{t+1})\}_{T_{upd}}$;
            \vspace{6pt}
            \State \underline{\textbf{\textit{Q-Value Function Update \& Data Sending:}}}
            \If{$t_{train} \geq T_{wup}$ \& the timer expires}
                \For{$k = 1, \dots, K$}
                    \State Sample mini-batch $h^{k}$ of size $h_{size}$ from $H$;
                    \State Update $\psi$ using $h^{k}$ and corresponding $\mathbf{\bar{a}}^{(i)}_{t+1}$,\Statex ~~~~~~~~~~~~~~~~~~~~$\log(\pi_{\bar{\boldsymbol{\uptheta}}}(\mathbf{\bar{a}}_{t+1}|\mathbf{o}_{t+1}))$ according to Eq. \eqref{critic_loss_alg};
                    \State Update target critic: $\bar{\psi} = \kappa\bar{\psi} + (1-\kappa)\psi$;
                \EndFor
                \For{$i \in \mathcal{I}_p$}\Comment{In parallel.}
                    \For{$k = 1, \dots, K$}\Comment{All together.}
                        \State \textbf{\textit{Compute}} $Q^{\psi,(i)}(\mathbf{o}_t,\mathbf{\hat{a}}_t)$ and $b(\mathbf{o}_t,\mathbf{\hat{a}}^{(\backslash i)}_t)$ based
                        \Statex ~~~~~~~~~~~~~~~~~~~~~~~~on $\mathbf{o}_t\in h^k$ and $\mathbf{\hat{a}}_t$; 
                        \State \textbf{\textit{Send}} $o^{(i)}_t \in h^{k}$,
                        and corresponding values
                        \Statex ~~~~~~~~~~~~~~~~~~~~~~~ of $Q^{\psi,(i)}(\mathbf{o}_t,\mathbf{\hat{a}}_t)$ and $b(\mathbf{o}_t,\mathbf{\hat{a}}^{(\backslash i)}_t)$ to agent $i$;
                    \EndFor
                \EndFor
            \EndIf

        \EndFor
    \end{algorithmic}
\end{algorithm}

\begin{algorithm}[t]
    \small
    \caption{\small\textit{Learning Procedures for Each Agent $i$}}
    \label{learning_alg_each_agent}
    Parameters: $T_{train}$, $T_{upd}$, $T_{wup}$, $K$, $\kappa$.
    \begin{algorithmic}[1]
        \State Initialize $\theta^{(i)}$;
        \For{$t_{train}=1, \dots, T_{train}$}
            \vspace{6pt}
            \State \underline{\textbf{\textit{Data Arrival Detection:}}}
            \For{$k = 1, \dots, K$}
                \State \textbf{\textit{Read}} $o^{(i)}_t\in h^{k}$,
                corresponding values of
                \Statex ~~~~~~~~~~~~$Q^{\psi,(i)}(\mathbf{o}_t,\mathbf{\hat{a}}_t)$ and $b(\mathbf{o}_t,\mathbf{\hat{a}}^{(\backslash i)}_t)$ from central entity;
            \EndFor
            \vspace{6pt}
            \State \underline{\textbf{\textit{Policy Update \& Data Sending:}}}
            \If{$t_{train} \geq T_{wup}$ \& the timer expires}
                \For{$k = 1, \dots, K$}
                    \State Update $\theta^{(i)}$ using $o^{(i)}_t$, $Q^{\psi,(i)}(\mathbf{o}_t,\mathbf{\hat{a}}_t)$, $b(\mathbf{o}_t,\mathbf{\hat{a}}^{(\backslash i)}_t)$ 
                    \Statex ~~~~~~~~~~~~~~~~~according to Eq. \eqref{policy_gradient_alg};
                    \State Update target policy: $\bar{\theta}_i = \kappa\bar{\theta}_i + (1-\kappa)\theta_i$;
                \EndFor
                \vspace{4pt}
                \State Interact with the env. using newly updated policy;
                \vspace{4pt}
                \State \textbf{\textit{Compute}} $\hat{a}_t^{(i)}$ based on current policy $\pi_{\theta^{(i)}}$ and $\bar{a}_{t+1}^{(i)}$, \Statex ~~~~~~~~~~~~$\log(\pi_{\bar{\theta}^{(i)}}(\bar{a}_{t+1}^{(i)}|o_{t+1}^{(i)}))$ based on target policy $\pi_{\bar{\theta}^{(i)}}$;
                \State \textbf{\textit{Send}} latest $T_{upd}$ tuples
                  $\{(o^{(i)}_t,a^{(i)}_t,r^{(i)}_t,o_{t+1}^{(i)})\}_{T_{upd}}$
                  \Statex ~~~~~~~~~~~~and the corresponding values $\hat{a}_t^{(i)}$, $\bar{a}_{t+1}^{(i)}$, 
                  \Statex ~~~~~~~~~~~~$\log(\pi_{\bar{\theta}^{(i)}}(\bar{a}_{t+1}^{(i)}|o_{t+1}^{(i)}))$ to the central entity;
            \EndIf
        \EndFor
    \end{algorithmic}
\end{algorithm}

The amount of information sent by each remote agent (IAB-node array panel) $i$ to the central entity (IAB-donor), shown by blue arrows \ding{172} in Fig.~\ref{fig:com_coop_combined}, is dominated by the set of tuples $Tup^{(i)}=\{(o^{(i)}_t, a^{(i)}_t, r^{(i)}_t, o^{(i)}_{t+1})\}_{T_{upd}}$. The size of each tuple is mainly determined by the size of an observation $o^{(i)}_t$, which consists of $N_s + N_s\cdot A + N^{ch}_{i}\cdot L$ bits for FD case and $N_s + N_s\cdot A + N^{ch}_{i}\cdot(L + B)$ bits for HD case, where $N_s$ is given by the UE presence ($I_{i,s}^{pres}$) bitmap, $A$ is the number of bits used to record the average attenuation ($A^{block}_{i,s}$) caused by obstacles in a sector, $L$ and $B$ are respectively the numbers of bits used to indicate the buffer level and the amount of data bits delivered by a child IAB-node, and $N^{ch}_i$ is the number of child IAB-nodes connected to $i$, which is typically small (e.g., $2$-$4$).
In addition, each agent sends, together with each tuple $tup^{(i)}=(o^{(i)}_t, a^{(i)}_t, r^{(i)}_t, o^{(i)}_{t+1})\in Tup^{(i)}$, the following variables:
\begin{itemize}
    \item action $\hat{a}^{(i)}_t$ that it would play in front of current observation $o^{(i)}_t$ in $tup^{(i)}$, selected according to the current policy function $\pi_{\theta^{(i)}}(\hat{a}^{(i)}_t | o^{(i)}_t)$;
    \item action $\bar{a}^{(i)}_{t+1}$ that it would play in front of the next observation $o^{(i)}_{t+1}$ in $tup^{(i)}$, selected according to its target policy function $\pi_{\bar{\theta}^{(i)}}(\bar{a}^{(i)}_{t+1} | o^{(i)}_{t+1})$; 
    \item the value of $\log(\pi_{\bar{\theta}^{(i)}}(\bar{a}^{(i)}_{t+1}|o^{(i)}_{t+1}))$, conditional to $o^{(i)}_{t+1}$ in $tup^{(i)}$ and according to the target policy function $\pi_{\bar{\theta}^{(i)}}(\bar{a}^{(i)}_{t+1}|o^{(i)}_{t+1})$. 
\end{itemize}
Actions $\bar{a}^{(i)}_{t+1}$ ($i\in\mathcal{I}_p$) and the values of $\log(\pi_{\bar{\theta}^{(i)}}(\bar{a}^{(i)}_{t+1}|o^{(i)}_{t+1}))$ ($i\in\mathcal{I}_p$) are used in the centralized action-value function updates according to Eqs.~\eqref{critic_loss_alg} and \eqref{y_i_alg}, while $\hat{a}^{(i)}_t$ is used at the central entity to compute $Q^{\psi,(i)}(\mathbf{o}_t,\mathbf{\hat{a}}_t)$ and $b(\mathbf{o}_t,\mathbf{\hat{a}}^{(\backslash i)}_t)$ that will be redistributed to all the agents to perform policy updates.

\vspace{-10pt}
\subsection{Training Details for Distributed Agents}
Once the $Q^{\psi,(i)}(\mathbf{o}_t,\mathbf{a}_t)$ ($i\in\mathcal{I}_p$) functions are updated, the central entity immediately sends to each agent $i$ the information necessary to update its policy (shown by green arrows \ding{173} in Fig.~\ref{fig:com_coop_combined}), which consists of:
\begin{itemize}
    \item observations $\{o^{(i)}_t\}_{h_{size}}$ extracted from tuples in $h$;
    \item values of $Q^{\psi,(i)}(\mathbf{o}_t,\mathbf{\hat{a}}_t)$ and $b(\mathbf{o}_t,\mathbf{\hat{a}}^{(\backslash i)}_t)$.
\end{itemize}
Based on this information, each agent $i$ performs gradient ascent to update $\theta^{(i)}$, as in Algorithm \ref{learning_alg_each_agent} [Line 9], and obtains a new $\pi_{\theta^{(i)}}(a^{(i)}_t|o^{(i)}_t)$, which will be locally used in the next steps until a new updated policy is generated.
The parameters of the agent $i$'s policy are updated as follows:
\begin{align}
    &\nabla_{\theta^{(i)}}J(\pi_{\boldsymbol{\uptheta}}) = \nabla_{\theta^{(i)}} \log( \pi_{\theta^{(i)}}(\hat{a}^{(i)}_t|o^{(i)}_t) )\cdot\nonumber\\
    &\left(-\tau\log( \pi_{\theta^{(i)}}(\hat{a}^{(i)}_t|o^{(i)}_t) )+
    Q^{\psi,(i)}(\mathbf{o}_t,\mathbf{\hat{a}}_t) - b(\mathbf{o}_t,\mathbf{\hat{a}}^{(\backslash i)}_t)\right).
    \label{policy_gradient_alg}
\end{align}
Note that action $\hat{a}^{(i)}_t$ is the same as the one sent to the central entity. Indeed, it is generated by the agents to be sent to the central entity and then stored to be used for computing Eq.~\ref{policy_gradient_alg}. As a final task, each agent updates its target policy by using a moving average.

It is worth mentioning that the policy updates can be performed in parallel to normal network operations. When a concurrent policy update takes place, the actions for normal network operations are selected according to the latest version of the policy function, which is being updated and will be improved at the end of the current policy update.
Note that only the experience data collected when applying fresh new policies are sent to the central entity for the Q-value function updates, as illustrated by the orange dashed arrow in Fig. \ref{fig:com_coop_combined}(b), while the other interactions are used just to keep the IAB network continuously active.

\vspace{-5pt}
\section{Numerical Results}
\label{experiments}
In this section, we evaluate the performance of the proposed MARL-based resource allocation approach on several instances containing IAB-nodes working in FD or HD mode, mobile UEs, and random link failures caused by mobile obstacles. Every value shown in the figures of this section is the result of an average over 10 random instances.

\vspace{-11pt}
\subsection{Scenario Settings}
In line with 3GPP NR IAB simulation guidelines \cite{3gpp15study}, we consider a $300$m$\times300$m service area where 1 IAB-donor is located at the left-side midpoint and 4 IAB-nodes are randomly deployed in the area. Fig.~\ref{fig:example_simulation_scenarios} shows two examples of backhaul deployment of IAB network scenarios.
A set of 30 UEs move around in the area, with random initial positions and directions. The considered heights of the IAB-donor, IAB-nodes and UEs are $25$m, $6$m and $1.5$m, respectively.
Both the IAB-donor and IAB-nodes are equipped with $N_p=4$ antenna panels, each of which contains $8\times6$ elements and manages $N_s=5$ sectors.
The transmission power of each panel at the IAB-donor and IAB-nodes is respectively $29.3$ dBm and $20.3$ dBm. The receiver noises at the IAB-nodes and UEs are $-84.023$ dBm and $-82.023$ dBm, respectively. The azimuth HPBWs for the IAB-donor and IAB-nodes are $\pi/36$ and $\pi/12$, and the elevation HPBW is $\pi/4$ for both. The SINR thresholds and rates considered for the access and backhaul links are those indicated in MCS Index Table 3 for PDSCH in the 3GPP NR specification \cite{3gpp17PhysicalData}. Finally, one frame consists of $80$ slots, each with a duration of $\delta=125\mu$s, which correspond to the in-band IAB at $28$ GHz, $400$ MHz bandwidth, NR Numerology \#$3$ ($120$ kHz subcarrier spacing).

\textbf{User mobility:}
UEs move in the playground according to a Random Waypoint model \cite{bai2004survey}. Specifically, a UE randomly selects a direction in the angular range $\xi \in [-180^\circ, 180^\circ]$ from the current direction. Then it travels along the selected direction with a constant speed uniformly chosen within the range $[2, 20]$m/s (or $[20,60]$m/s in the extended analysis). Speeds and directions of different UEs are selected independently and randomly. After moving for $t_{m}\in [2, 6]$s, a UE pauses for an interval $t_{p}\in [0, 1]$s before resuming. UEs bounce back when they reach the area boundary.

\textbf{Obstacles:}
We assess the performance of our approach according to two levels of obstacle densities in the area and refer to them as \emph{low obstacle density (LOD)} and \emph{high obstacle density (HOD)}. They are implemented by dropping in the simulated service area, respectively, $15$ and $60$ cylindrical obstacles with a radius of $2.5$m and a height of $2$m. They move at a speed of $2$m/s - $20$m/s following the same Random Waypoint model applied to UEs.

\begin{figure}
    \centering
    \includegraphics[width=0.8\linewidth]{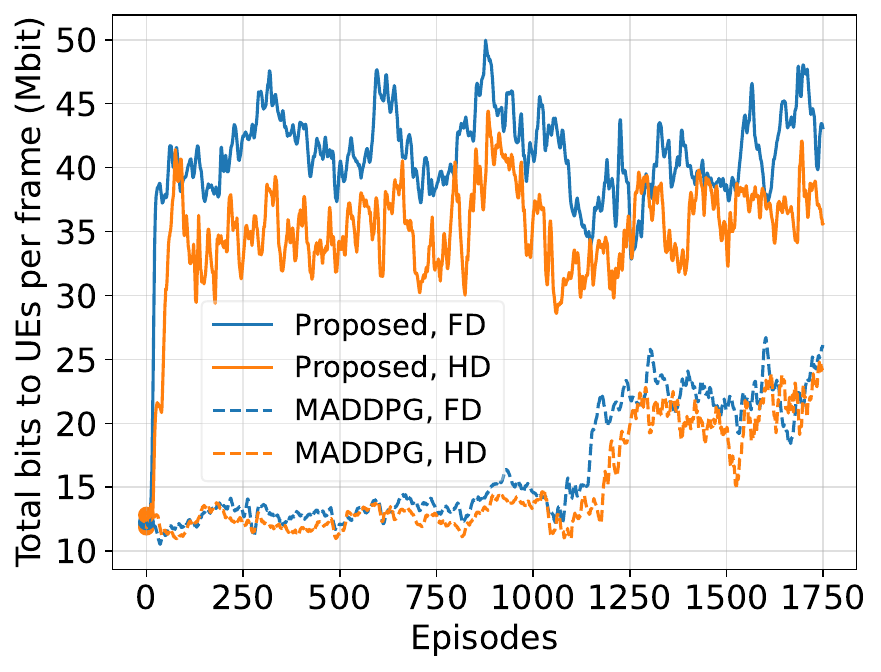}
    \caption{\small Training curves for FD and HD IAB networks (with user speeds in the range of 2-20m/s and under HOD condition).}
    \label{fig:training_curves_fd_hd}
    \vspace{-2mm}
\end{figure}

\begin{figure}[!t]
    \centering
    \includegraphics[width=0.48\linewidth]{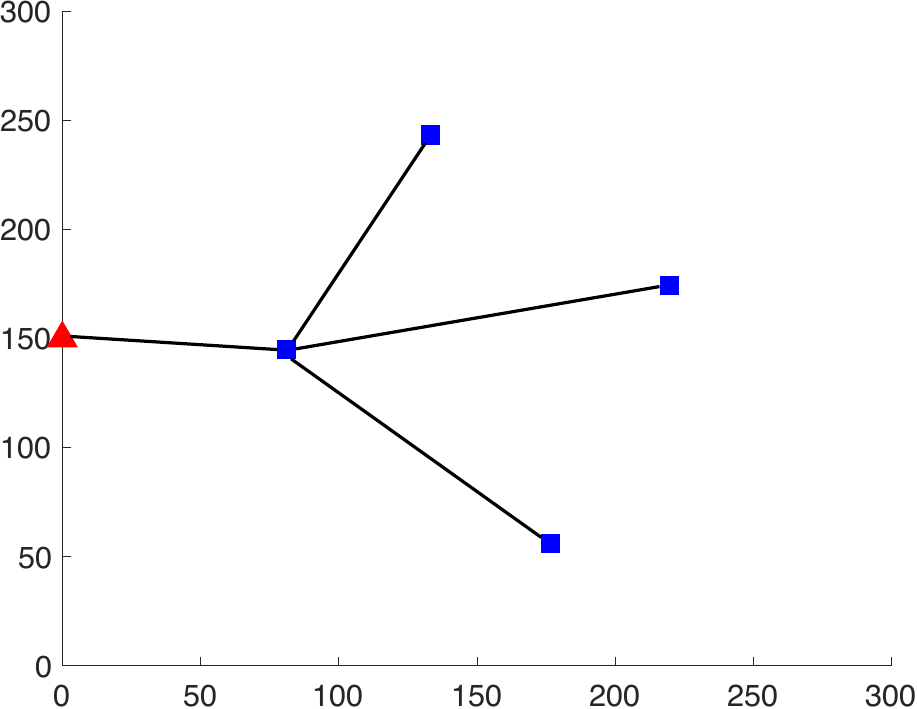}
    \includegraphics[width=0.48\linewidth]{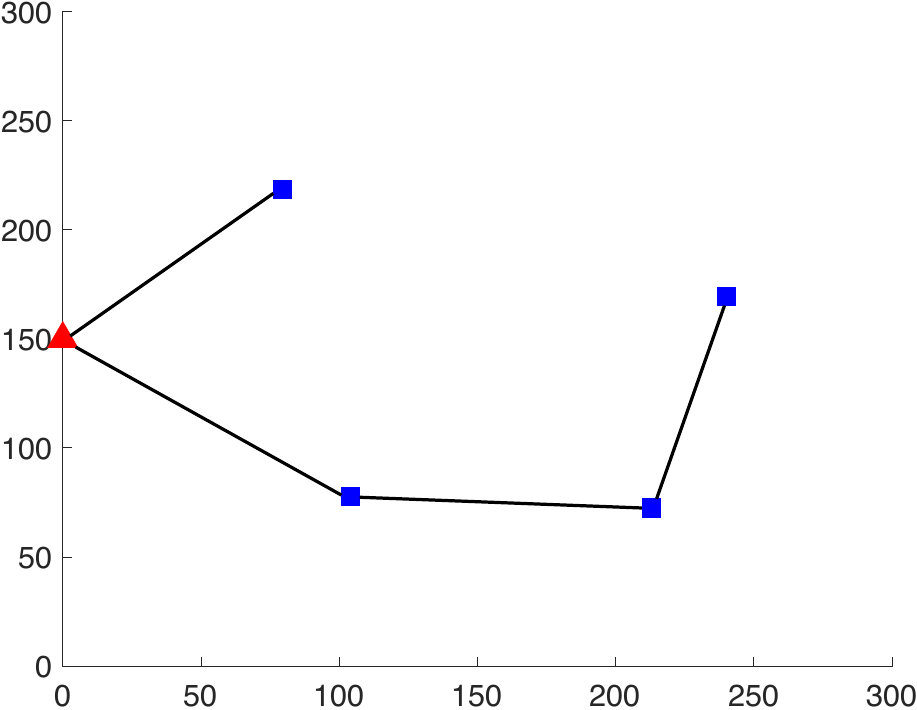}
    \caption{Example IAB network scenarios considered in the experiments (red triangle: IAB-donor, blue squares: IAB-nodes, black lines: backhaul links).}
    \label{fig:example_simulation_scenarios}
\end{figure}

\vspace{-12pt}
\subsection{MARL Model Settings}
We train each DNN model of our approach based on the experience data of $5000$ episodes, each of which consists of $T=80$ steps (slots).
This corresponds to a training period of $5000$ frames, thus a total time of $50$s. All the DNN models have a hidden dimension of $128$. We consider $K=4$ consecutive DNN updates, a data collection period of $T_{upd}=100$ steps, and a warm-up period of $T_{wup}=10$ episodes. The updates are performed via Adam optimizer with a learning rate of $0.001$ for both distributed policies and centralized critics. The weight $\rho_{BH}$ for backhaul transmission in the reward function is empirically set to $0.8$. The discount factor $\gamma $ is $0.99$ and the weight $\tau$ of the policy entropy is set to $0.01$. The weight vector $\bar{\psi}$ of the target critic network, similarly to $\bar{\theta}$ of the target actor network, is updated via $\bar{\psi} = \kappa\bar{\psi} + (1-\kappa)\psi$ with update rate $\kappa = 0.001$. The replay buffer $H$ has a maximum length of $10^6$ entries, and each update uses a mini-batch of $1024$ entries, randomly sampled from $H$.

The settings of the DNN architectures and training hyper-parameters for MADDPG baseline approach are the same accordingly.

Training curves expressing the total traffic volume delivered in a frame from the IAB-donor to all the UEs (which corresponds to the throughput objective of the resource optimization problem) are shown in Fig. \ref{fig:training_curves_fd_hd}, where IAB-nodes operate in FD and HD modes, respectively. To better appreciate the learning trend, we only show the first $1750$ episodes, which involve the key performance-improving period.
As we can see, compared with the MADDPG approach that shows apparent throughput boost after $1000$ episodes, the training process of the proposed approach shows an immediate throughput increase in both FD and HD cases. Indeed, blue and orange curves can reach high values within $100$ episodes, showing the models can learn fast from the experience. Nevertheless, we train both approaches up to $5000$ episodes to let them accumulate sufficient experience and eventually obtain a stable performance in any situation. Moreover, the proposed approach can achieve almost double throughput of the MADDPG, which is consistent with
Figs.~\ref{fig:bars_fd}, \ref{fig:bars_hd}, \ref{fig:cdf_datarate}.

\vspace{-10pt}
\subsection{Performance Analysis}

\begin{figure*}
\centering
    \begin{subfigure}[b]{0.32\textwidth}
            \centering
            \includegraphics[width=\linewidth]{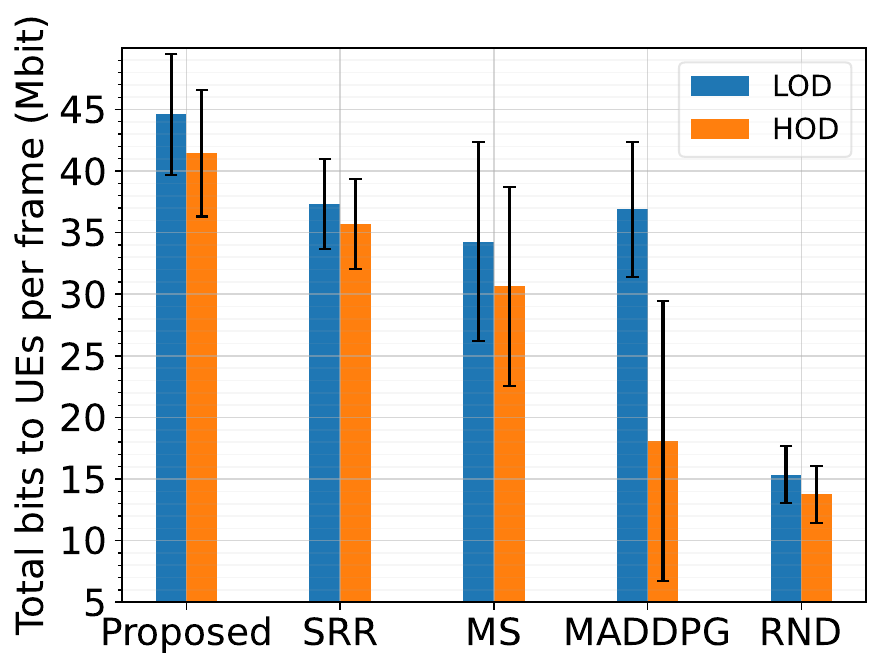}
            \caption{\small Avg. bit volume delivered per frame.}
            \label{total_bits_fd}
    \end{subfigure}
    \begin{subfigure}[b]{0.32\textwidth}
            \centering
            \includegraphics[width=\linewidth]{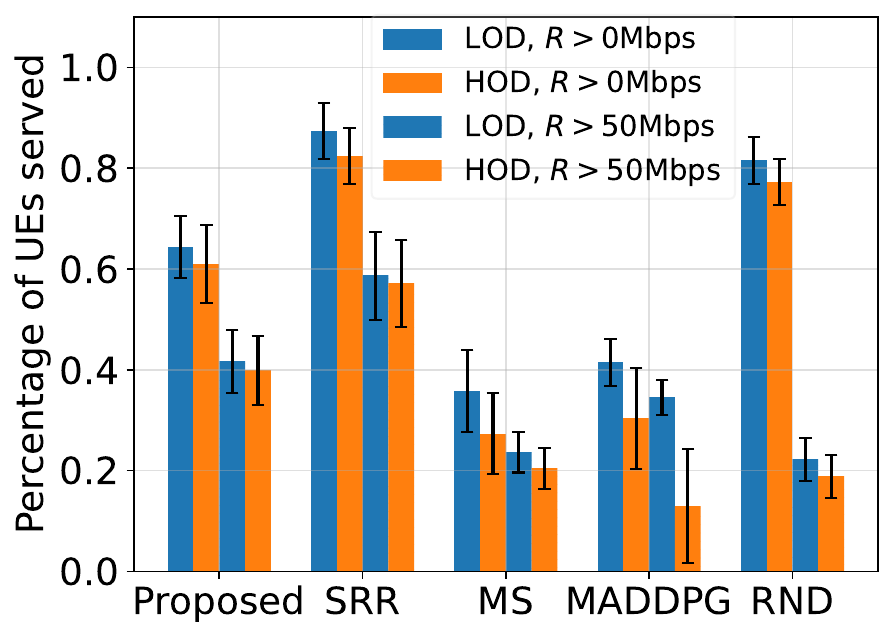}
            \caption{\small Avg. percent of served UEs per frame.}
            \label{UE_percent_fd}
    \end{subfigure}
    \begin{subfigure}[b]{0.32\textwidth}
            \centering
            \includegraphics[width=\linewidth]{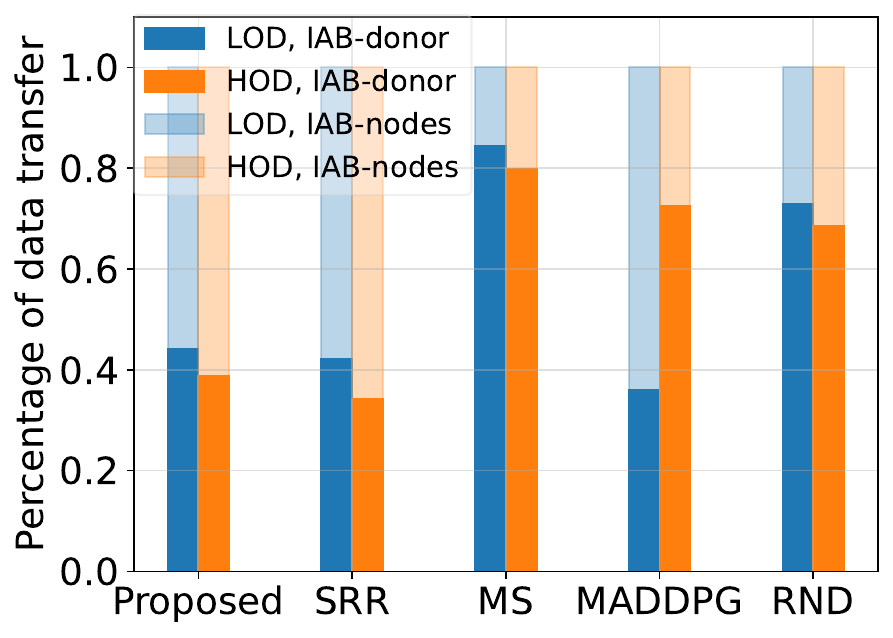}
            \caption{\small Avg. traffic volume split per frame.}
            \label{BSRN_percent_fd}
    \end{subfigure}
    \caption{\small Performance comparison of the five schemes in FD IAB networks.}
    \label{fig:bars_fd}
    \vspace{-4mm}
\end{figure*}

\begin{figure*}
\centering
    \begin{subfigure}[b]{0.32\textwidth}
            \centering
            \includegraphics[width=\linewidth]{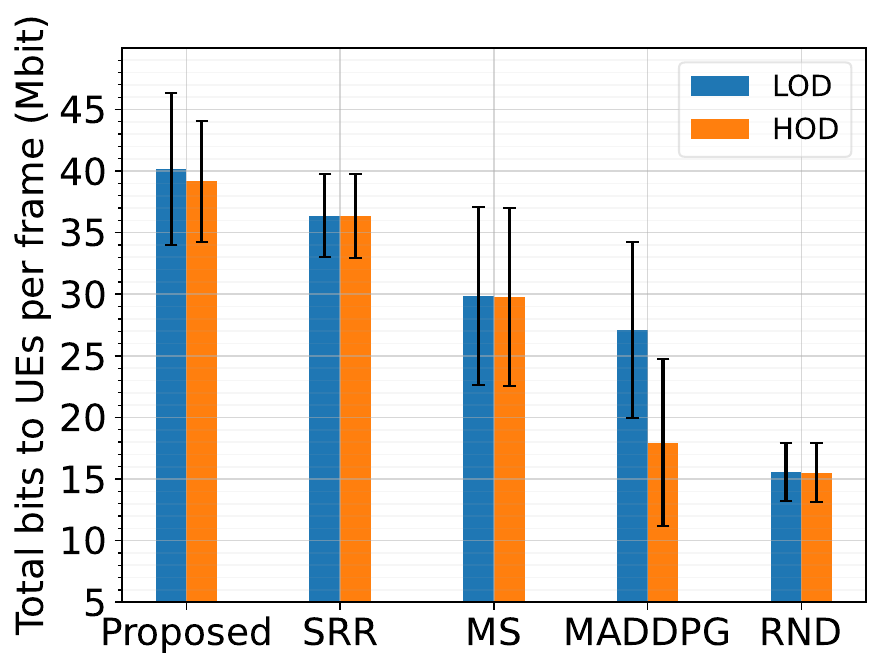}
            \caption{\small Avg. bit volume delivered per frame.}
            \label{total_bits_hd}
    \end{subfigure}
    \begin{subfigure}[b]{0.32\textwidth}
            \centering
            \includegraphics[width=\linewidth]{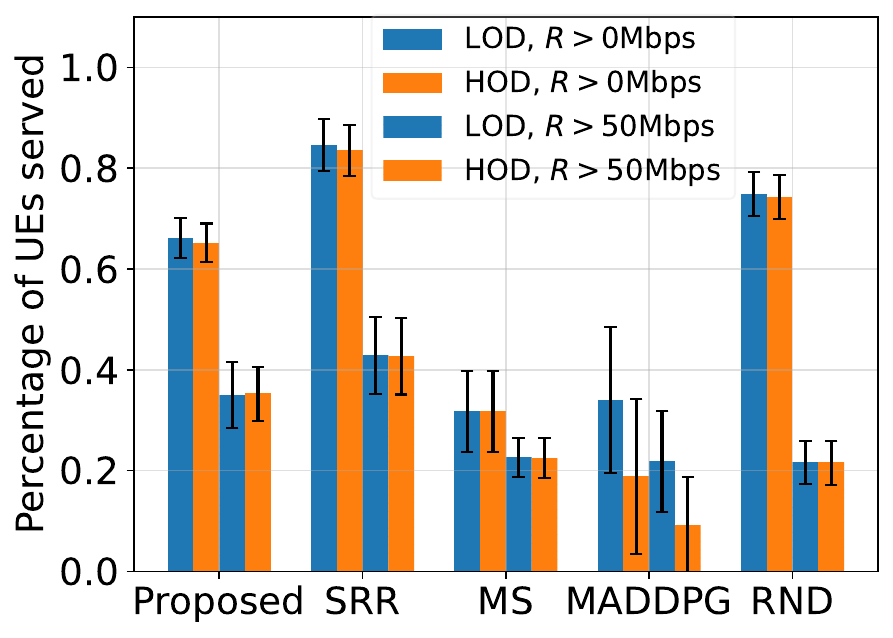}
            \caption{\small Avg. percent of served UEs per frame.}
            \label{UE_percent_hd}
    \end{subfigure}
    \begin{subfigure}[b]{0.32\textwidth}
            \centering
            \includegraphics[width=\linewidth]{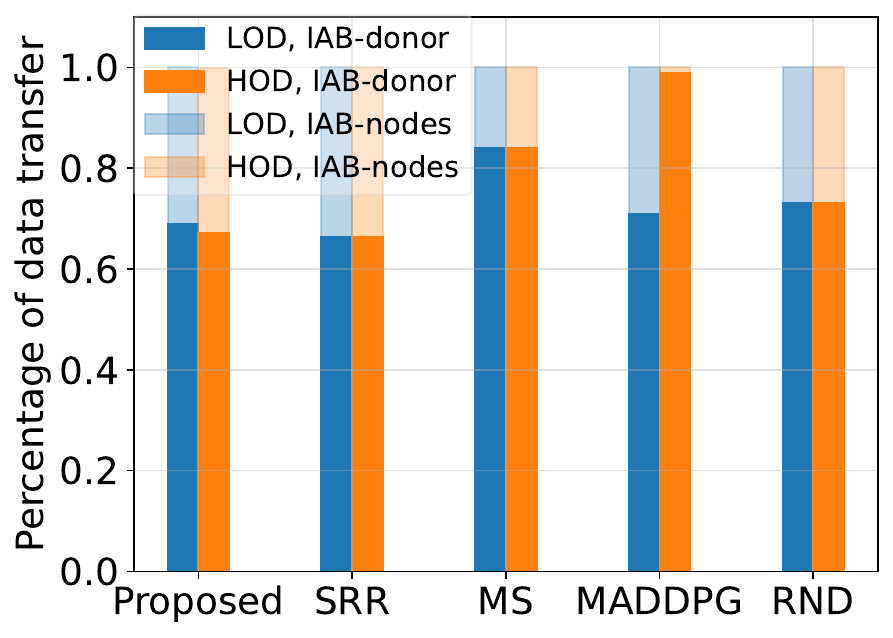}
            \caption{\small Avg. traffic volume split per frame.}
            \label{BSRN_percent_hd}
    \end{subfigure}
    \caption{\small Performance comparison of the five schemes in HD IAB networks.}
    \label{fig:bars_hd}
    \vspace{-4mm}
\end{figure*}

\begin{figure*}
\centering
    \begin{subfigure}[b]{0.38\textwidth}
		\centering
   		\includegraphics[width=\linewidth]{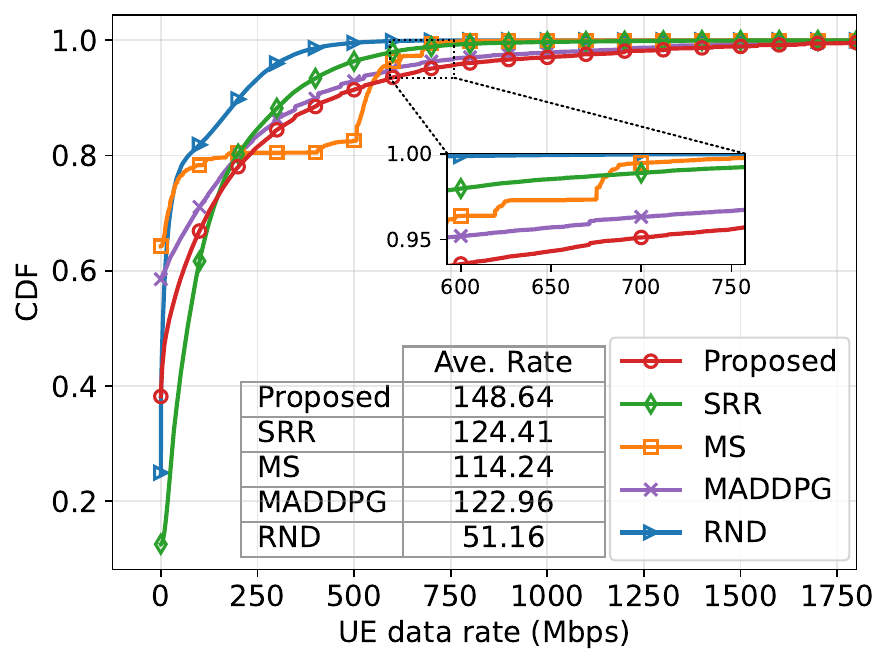}
		\caption{\small UE data rate in the FD case.}
    	\label{fig:cdf_fd}
	\end{subfigure}
	\begin{subfigure}[b]{0.38\textwidth}
   		\centering
  		\includegraphics[width=\linewidth]{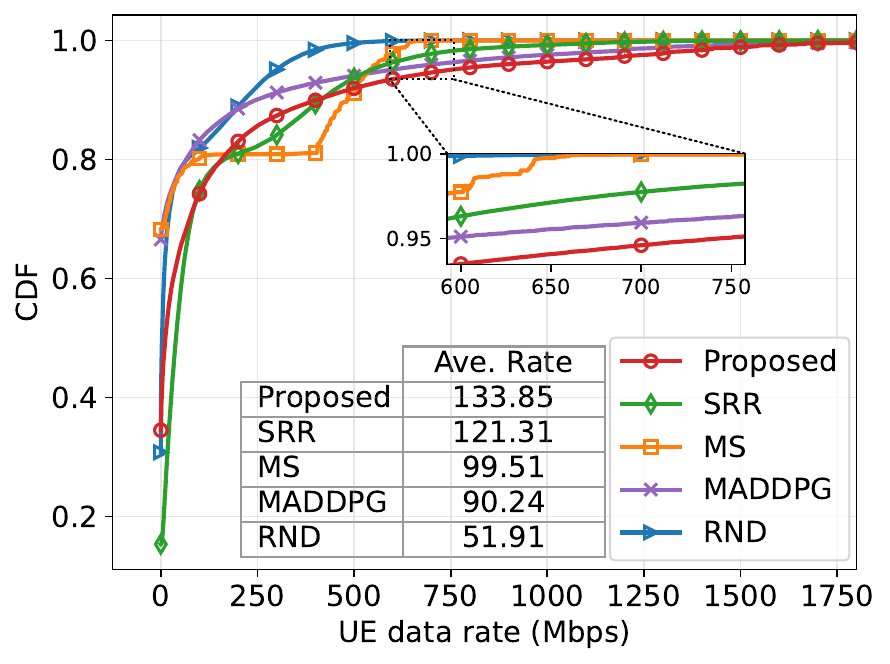}
   		\caption{\small UE data rate in the HD case. }
  		\label{fig:cdf_hd}
	\end{subfigure}
	\caption{\small CDF of the data rate achieved by each UE in a frame, under the LOD condition.}
	\label{fig:cdf_datarate}
        \vspace{-5mm}
\end{figure*}

\begin{figure*}
\centering
    \begin{subfigure}[b]{0.38\textwidth}
   		\centering
  		\includegraphics[width=\linewidth]{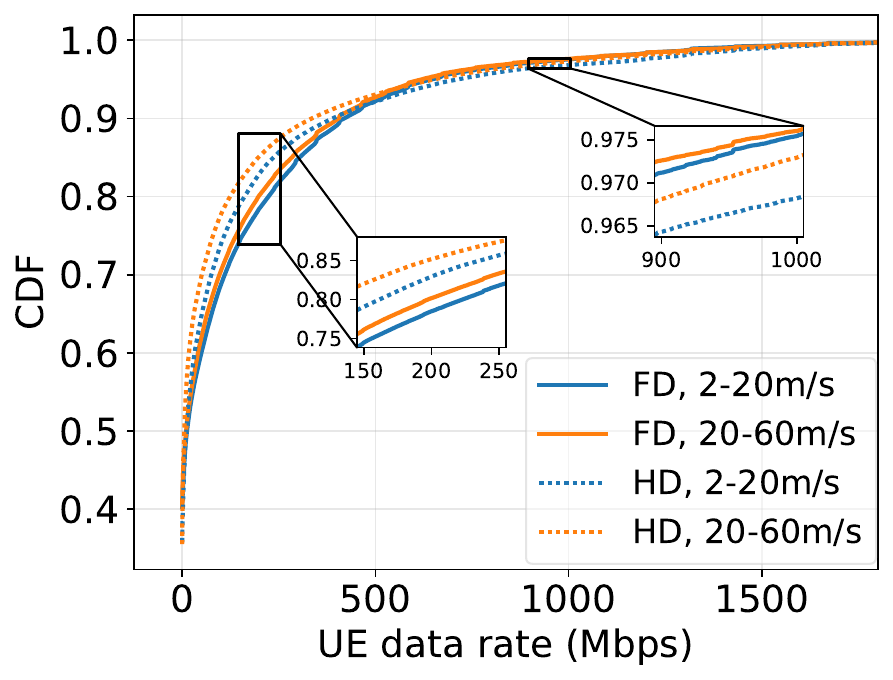}
   		\caption{\small CDF of UE data rate.}
  		\label{fig:diff_speed_count_for_every_ue}
	\end{subfigure}
    \begin{subfigure}[b]{0.38\textwidth}
		\centering
   		\includegraphics[width=\linewidth]{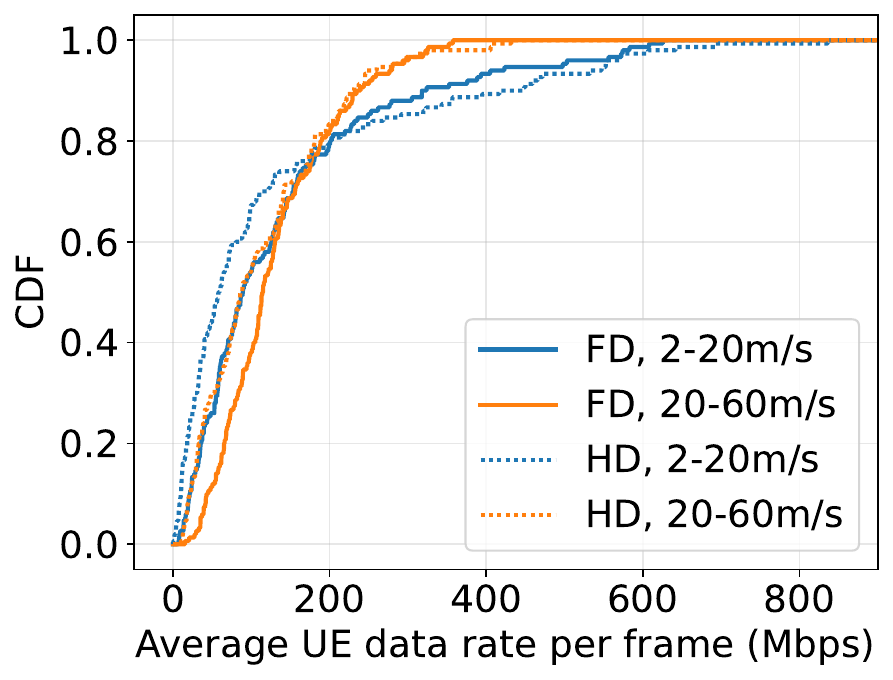}
		\caption{\small CDF of average UE data rate per frame.}
    	\label{fig:diff_speed_average_per_frame}
    \end{subfigure}
	\caption{\small CDFs of the UE data rates achieved at different UE speeds and duplex modes.}
	\label{fig:cdf_diff_speed}
    \vspace{-6mm}
\end{figure*}

We compare the proposed MARL-based approach (referred to as \textit{Proposed} in the following) against four representative schemes:
\begin{itemize}
	\item Super Round-Robin scheme (referred to as \textit{SRR}):\\
    A common scheduling scheme when dealing with wireless access transmissions, where each panel of the IAB-donor and IAB-nodes iteratively serves slot-by-slot every UE under its coverage in round-robin fashion. In this scheme, all the IAB-nodes are assumed to always have data bits to deliver, which is guaranteed by the automatic IAB-node's buffer refill from its parent node when its number of bits drops below a certain threshold. This is an ideal behavior (the reason why it is named ``super"); indeed, a proper refilling strategy must be designed. \textit{SRR}'s performance provides an upper bound on the performance of any real round-robin implementation.
	\item Multi-Slot algorithm (referred to as \textit{MS}):\\
 A heuristic algorithm proposed in \cite{saad2019millimeter} to perform link scheduling. This algorithm generates a sequence of link sets, each of which contains a group of links that can be simultaneously activated in a slot satisfying the SINR conditions required by activated MCSs. Based on this algorithm, we periodically generate the compatible link sets (e.g., every frame) or timely regenerate the link sets every time the network layout undergoes a change due to mobile users, and iteratively apply them in a sequential order, slot by slot, till the next link set generation.
        
        \item MADDPG-based scheme (referred to as \textit{MADDPG}):\\ A scheme obtained through training the agents formulated in Sec~\ref{method} based on MADDPG algorithm \cite{maddpg}. 
        
        \item Random scheme (referred to as \textit{RND}):\\
        A random scheduling scheme where each panel randomly picks an action from its candidate action set.
\end{itemize}

We begin the analysis for both FD and HD cases by considering typical UE urban speeds, in the range of $2$m/s - $20$m/s, to assess the impact of different obstacle densities on the performance. Results are shown in Figs. \ref{fig:bars_fd}-\ref{fig:cdf_datarate}. Then, we extend our analysis to consider different speed ranges, as reported in Fig. \ref{fig:cdf_diff_speed}. We refer readers to Appendix Sec. B for a complexity and scalability analysis of the system.

\textbf{Traffic Volume:}
Figs. \ref{total_bits_fd} and \ref{total_bits_hd} show the average overall traffic volume delivered to UEs per frame, respectively in the FD case and in the HD case, considering both the LOD and HOD blockage situations.

Compared to \textit{SRR}, which assumes an ideal refilling strategy for IAB-nodes' buffers, the \textit{Proposed} can achieve an improvement of $20\%$ in the FD case and $10\%$ in the HD case. This is because the \textit{Proposed} can coordinate the interference among links and adapt its decisions to UE mobility and obstacle obstructions. The reduced gain in the HD mode is mainly due to the limited degrees of freedom caused by HD constraints at IAB-nodes.
Indeed, on one side, an IAB-node's buffer needs to be refilled through backhaul transmissions in order not to bottleneck downstream access transmissions, on the other, backhaul transmissions to the IAB-node preclude its access transmissions due to HD constraints.
In practice, these HD limitations prevent smart resource allocation schemes from fully exploiting all available wireless resources.
Nevertheless, in the HD case, the \textit{Proposed} approach can still achieve an advantage of almost $500$Mbps over \textit{SRR}.

Moreover, the \textit{Proposed} outperforms \textit{MS} by $30\%$ in FD case and $35\%$ in HD case. Although \textit{MS} provides a near-optimal interference coordination and its compatible link sets are ideally regenerated whenever it is needed, this cannot compensate the \textit{Proposed}'s advantages, which timely recharges IAB-nodes' buffers and on-the-fly adapts to obstacle blockages. In addition, we can observe an interesting aspect: \textit{SRR} performing better than \textit{MS} demonstrates how a good IAB-nodes' buffer management has a larger impact on the data transfer than link interference coordination.

Furthermore, despite sharing the same RL components, the \textit{Proposed} outperforms the \textit{MADDPG} by around $20\%$ and $60\%$ in LOD and HOD conditions for FD networks, and around $50\%$ and $55\%$ for HD networks, respectively. This primarily stems from the distinct model architectures, especially the attention-based central critics, and training procedures of the \textit{Proposed} and \textit{MADDPG}.

Finally, according to the error bars at the tips of grouped bars, which represent the standard deviations of the throughput delivered to UEs, \textit{RND} and \textit{SRR} exhibit the smallest variance, while the \textit{MS} and \textit{MADDPG} show the largest variance. The variance of the \textit{Proposed} approach is reasonably small, which demonstrates a stable performance across different network scenario instances.

\textbf{Coverage:}
Figs. \ref{UE_percent_fd} and \ref{UE_percent_hd} indicate the average percentage of UEs served per frame. We adopt two definitions of served UEs: in the first one, a UE is declared as served if it experiments a data rate in a frame larger than $0$Mbps ($R>0$Mbps in the figures), while the second definition introduces a minimum data rate threshold of $50$Mbps ($R>50$Mbps in figures). Similarly to the previous traffic volume figures, results in both the LOD and HOD conditions are shown.

Some evident aspects emerge from the figures. Considering the minimum data rate of $0$Mbps, both \textit{SRR} and \textit{RND} show higher percentages of $(R>0)$-served UEs than the \textit{Proposed} scheme. This is a consequence of the throughput-vs.-fairness trade-off. While \textit{SRR} and \textit{RND} reach all UEs with the same probability thus providing the best fairness, the \textit{Proposed} tends to preferably serve those UEs that can bring the best overall throughput. However, when considering UEs that are served with at least $50$Mbps, the gap between \textit{SRR} and the \textit{Proposed} remarkably reduces and the \textit{Proposed} even outperfoms \textit{RND}. This confirms that the service percentages guaranteed by \textit{SRR} and \textit{RND} are mainly driven by UEs with very-low data rates. This further demonstrates that the proposed resource allocation scheme can very effectively deal with such complex network scenarios.

\textbf{Backhaul Load:}
Figs. \ref{BSRN_percent_fd} and \ref{BSRN_percent_hd} provide an insight into the average fraction of the traffic delivered to UEs through the IAB wireless backhaul per frame. The upper translucent part of each bar represents the average percentage of the data volume received by UEs from IAB-nodes via multi-hops, while the lower opaque part reports the complementary percentage of the traffic directly received from the IAB-donor. We can see that the \textit{Proposed} aggressively resorts to backhaul IAB-nodes when operating in FD mode, even if the larger panels and the higher transmission power of the IAB-donor may lead UEs to directly connect to it. In the HD case, all the considered schemes reduce the load of the wireless backhaul. Indeed, HD IAB-nodes are less effective in relaying traffic flows, thus limiting wireless link transmission concurrency, especially in the IAB tree topology.

\textbf{CDF of Per-UE Data Rate:}
Figs. \ref{fig:cdf_fd} and \ref{fig:cdf_hd} compare the performance of the five schemes in terms of per-UE data rate cumulative distribution function (CDF). As the CDF curves show very similar trends under LOD and HOD conditions, we only show LOD curves. We measure the per-UE data rate frame by frame, whose values are used to compute the CDF.

As we can see from the upper right corner of the figures, the maximum rate achieved by the \textit{Proposed} is near 1800 Mbps in both FD and HD cases, which remarkably exceeds the other four schemes.
This means that the problem faced is not trivial and only a careful link scheduling scheme can allow a good performance. In addition, this further proves that the \textit{Proposed} can discover the most effective strategy to increase the overall throughput.

Moreover, it is evident that the order of the five schemes to serve high per-UE data rate is: the \textit{Proposed}, \textit{MADDPG}, \textit{SRR}, \textit{MS} and \textit{RND}. This shows an interesting point that in order to maximize total throughput, learning-based approaches (\textit{Proposed} and \textit{MADDPG}) tend to pursue large per-UE data rate rather than serve a large number of UEs with average per-UE data rate.

The CDF values on the leftmost side indicate the percentage of users that cannot be served. This information has been better described through the solid bars ($R>0$Mbps) in Figs. \ref{UE_percent_fd} and \ref{UE_percent_hd}, however, here we can see how \textit{SRR} and \textit{RND} schemes show the highest probabilities for small rate values, because they do not tend to select the best UEs to maximize the overall throughput, but rather to reach all UEs with the same probability, although with a small throughput.

\textbf{UE Speed Sensitivity:}
Fig. \ref{fig:cdf_diff_speed} shows the performance of the \textit{Proposed} scheme over different speed ranges in both FD and HD cases. As the plots for the LOD and HOD cases are very similar, we only show the one of the HOD case. In particular, we observe the per-UE data rate CDF from two perspectives.
On one hand, as shown in Fig. \ref{fig:diff_speed_count_for_every_ue}, we adopt the same approach as in Fig. \ref{fig:cdf_datarate}, where we collect UE data rates frame by frame and compute the CDF based on all the collected rate values.
On the other, as shown in Fig. \ref{fig:diff_speed_average_per_frame}, we average the data rate of each UE over all the frames and compute the CDF based on per-UE data rate averages.

Fig. \ref{fig:diff_speed_count_for_every_ue} shows us that despite different speed ranges, the CDF curves corresponding to the same duplex mode are close. This implies that although the UE speed evidently affects the average UE data rate as indicated by Fig. \ref{fig:diff_speed_average_per_frame}, it has in practice a negligible impact on the distribution of the per-frame data rate across different UEs.

From Fig. \ref{fig:diff_speed_average_per_frame}, we can see that in both the FD and HD cases, increasing UE speeds reduces average UE data rates. This is reasonable because extremely fast-moving UEs can lead to more frequent and impactful network status changes.
However, even at the extremely high speeds of $[20,60]$m/s, which can be rarely seen in the urban scenarios where such IAB networks are envisioned, the impact on the performance is limited.

\vspace{-10pt}
\section{Conclusion}
\label{conclusion}
In this article, we have investigated the resource allocation problem in mmWave 5G IAB networks where user mobility and random obstructions caused by mobile obstacles produce strong network dynamics. Indeed, they generate short-lived access links and link-failure statistics that vary across different regions of the service area. Leveraging such scattered network behaviors, we have proposed an MARL-based approach that splits a combinatorial monolithic SARL problem, characterized by huge network state and action spaces, into smaller problems managed by different MARL agents.

Through the cooperation among MARL agents, the developed resource allocation approach can coordinate link interference and data caching on IAB-nodes, and capture network dynamics. We have designed different MARL setups for FD and HD node operations. Moreover, we have provided a learning framework considering potential feasibility issues (e.g., temporal dynamics) in real systems. The numerical results have shown that our MARL-based approach can achieve good throughput performance without significantly harming the network fairness.



\vspace{-10pt}
\renewcommand*{\bibfont}{\footnotesize}
\printbibliography

\vspace{-50pt}
\begin{IEEEbiography}[{\includegraphics[width=1in,height=1.25in,clip,keepaspectratio]{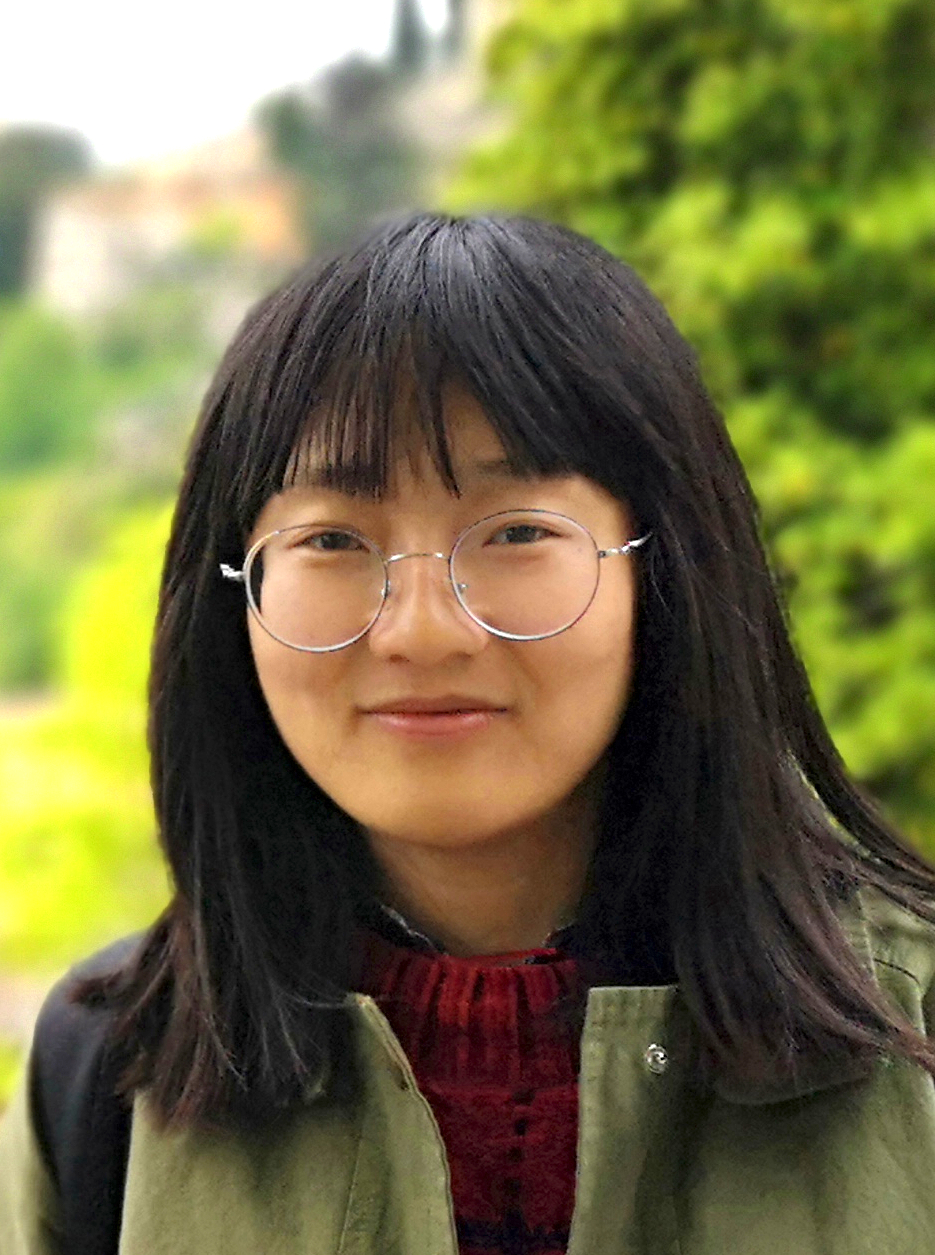}}]{Bibo Zhang}
received the B.S. degree in information engineering and the M.S. degree in electronics and communication engineering from Beijing University of Posts and Telecommunications, China, in 2015 and 2018, and the Ph.D. degree in information technology from Politecnico di Milano, Italy, in 2022. She is currently a Lecturer with the Ocean College, Jiangsu University of Science and Technology. Her research interests include resource management, wireless access networks, and artificial intelligence techniques.
\end{IEEEbiography}

\vspace{-50pt}

\begin{IEEEbiography}[{\includegraphics[width=1in,clip, keepaspectratio]{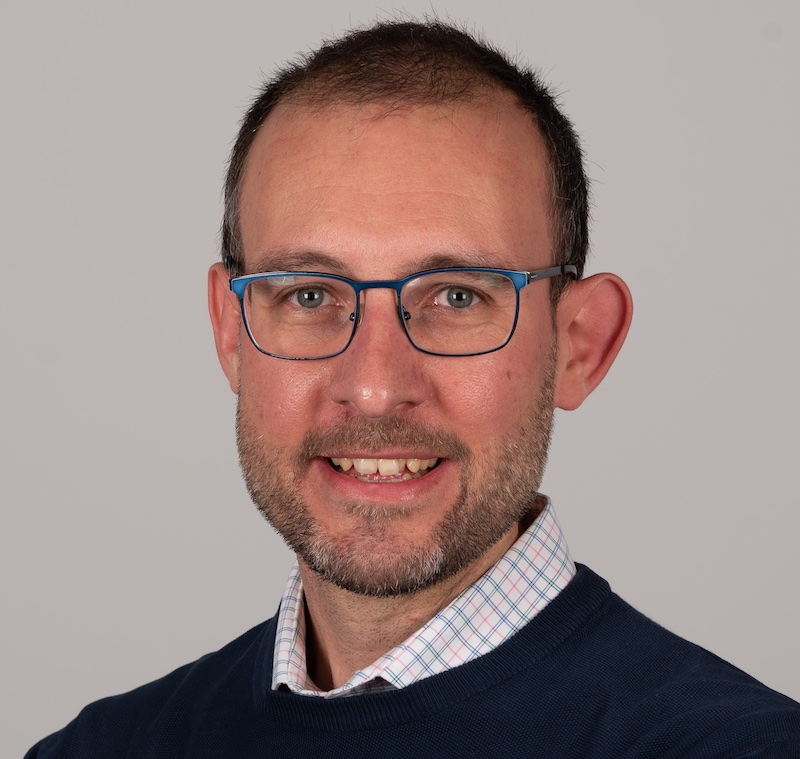}}]{Ilario Filippini}
received B.S. and M.S. degrees in Telecommunication Engineering and a Ph.D in Information Engineering from the Politecnico di Milano, in 2003, 2005, and 2009, respectively. He is currently an Associate Professor with the Dipartimento di Elettronica, Informazione e Bioingegneria, Politecnico di Milano. His research interests include planning, optimization, and game theoretical approaches applied to wired and wireless networks, performance evaluation and resource management in wireless access networks, and traffic management in software defined networks. He is an Associate Editor of \textit{Elsevier Computer Networks}.
\end{IEEEbiography}

\newpage
\vspace{-2mm}
\appendix
\vspace{-2mm}
\label{sec:appendix}
\subsection{Preliminaries of Single-Agent Actor and Critic}

RL was born as a tool to optimize decision-making and control, through the experience accumulated by an agent during sequential interactions with an environment, following a trial-and-error strategy. Specifically, at time step $t$, conditionally to the state $s_t \in \mathcal{S}$ of the environment, the agent selects an action $a_t \in \mathcal{A}$ according to its current policy $\pi$ and executes $a_t$ in the environment. At time step $t+1$, based on its reaction to $a_t$, the environment switches to state $s_{t+1}$ and gives a reward $r_t$ back to the agent. When states are partially-observable, an agent can only collect an observation $o_t \in \mathcal{O}$, which contains partial information of the global state $s_t$. Therefore, an action $a_t$ is selected based on the policy $\pi$ and conditionally to the current observation $o_t$. During the interactions, the agent adjusts the policy $\pi$ so as to maximize the long-term cumulative reward, namely the \emph{expected return} $\mathbb{E}_{\pi}[G_t] = \mathbb{E}[\sum^\infty_{k=t+1}\gamma^{k-t-1}r_k]$, where $\gamma$ is a discount factor controlling the importance of a future reward to the current utility.

A first type of learning approaches resort to an action-value function $Q(s_t,a_t)=\mathbb{E}_{\pi}[G_t|s_t, a_t]$ that corresponds to the expected return from state $s_t$, taking action $a_t$ and following policy $\pi$ afterwards.
An RL agent iteratively estimates this action-value function and selects at each step the action with the maximum function value in the current state. This is the fundamental idea of \textit{value-based} RL approaches. $Q^{\psi}(s_t,a_t)$ can be approximated as a $\psi$-parametric function of state and action, which can take the form of a deep neural network (DNN) with weights $\psi$. Parameters $\psi$ can be estimated via temporal-difference approaches by minimizing the regression loss $\mathcal{L}_{Q}(\psi)=\mathbb{E}_{(s_t,a_t,r_t,s_{t+1})\sim H}[(Q^{\psi}(s_t, a_t)-y)^2]$, where $y=r_t +\gamma\mathbb{E}_{a_{t+1}\sim\pi(s_{t+1})}[Q^{\bar{\psi}}(s_{t+1},a_{t+1})]$ is the updated return, $Q^{\bar{\psi}}$ is a moving average of past Q functions, and $H$ is a replay buffer storing past agent-environment interaction data tuples, including states, actions, and rewards.

A second family of learning techniques face the problem from a different perspective, and they are called \textit{policy-gradient} RL approaches. They see a policy as a function indicating the probability of selecting action $a_t$ in state $s_t$, parameterized with vector $\theta$, $\pi_{\theta}(a_t|s_t)=\Pr_{\theta}\{a_t|s_t\}$, which can be represented by a DNN as well, with $\theta$ as connection weights.
Parameters $\theta$ are updated by applying approximate gradient ascent to $\mathbb{E}[G_t]$, thus considering $\nabla_\theta\mathbb{E}[G_t]$, whose unbiased estimate is $\nabla_\theta \log \pi_{\theta}(a_t|s_t)G_t$. Further, $G_t$ can be approximated by its expectation $\mathbb{E}_{\pi}[G_t|s_t, a_t]$, which corresponds to the action-value function $Q^{\psi}(s_t, a_t)$. Finally, to reduce the estimate variance during updates, a state-dependent baseline $b(s_t)$ value is often subtracted from the unbiased estimate, which leads to the gradient $\nabla_\theta \log \pi_{\theta}(a_t|s_t)(Q^{\psi}(s_t, a_t)-b(s_t))$, where $Q^{\psi}(s_t, a_t)-b(s_t)$ is the \textit{advantage} of selecting action $a_t$ over other actions in state $s_t$.

The previous two paragraphs have briefly outlined the two main components of \textit{Actor-Critic} techniques \cite{mnih2016asynchronous}, which have emerged as one of the best-performing RL approaches. Indeed, they are derived from a policy-gradient approach, but incorporate the strengths of a value-based approach. In particular, the \emph{critic} part estimates the action-value function based on past interactions, thus generating $Q^{\psi}(s_t, a_t)$ values, while the \emph{actor} part updates the policy $\pi_{\theta}(a_t|s_t)$ according to the gradient direction, which in turn depends on the action-value function generated by the critic\footnote{Following the convention in Actor-Critic techniques, we will interchangeably refer to, respectively, critic function, action-value function or Q-value function as $Q$ and actor function or policy function as $\pi$ in the remainder of the article.}. Note that when considering partially-observable states, the policy becomes $\pi_{\theta}(a_t|o_t)$, which maps partial observation $o_t$ into a probability distribution over the action set. Similarly, the action-value function becomes $Q^{\psi}(o_t, a_t)$.

\subsection{Complexity and Scalability Analysis}

Let $U_{hid}$ denote the hidden dimension of each layer, then the total number of units in the central Q-value DNN is $U_Q=3U_{hid} + 5U_{hid} \cdot N$, where $N$ is the number of the agents. Specifically, $3U_{hid}$ counts the number of units in the shared attention part that consists of ``key", ``value" and ``agent selection", while $5U_{hid}$ corresponds to two embedding layers and $f^{(i)}$'s two layers where the first contains $2U_{hid}$ units and the second contains $U_{hid}$ units. Then, considering that each agent's policy network consists of 3 layers, the number of units in each agent's policy DNN is $3U_{hid}$, hence the total number of policy units is $U_A = 3U_{hid}\cdot N$. Therefore, the overall number of units is in the order of $\mathcal{O}(U_Q+U_A) = \mathcal{O}(3U_{hid}+8U_{hid}\cdot N)$, which approximates $\mathcal{O}(N)$, as $U_{hid}$ is a constant term potentially taking values of $\{8,16,32,64,128,256,512\}$ (e.g., $U_{hid}=128$ throughout the experiments in this article).
Based on the above analysis, the size of the DNN is only proportional to the number of IAB-nodes, which is usually smaller than 10, as indicated by 3GPP specifications.
From a different perspective, the action space of each agent only depends on the number of sectors of each antenna panel, which is constant in a specific environment, regardless of how many users exist in the system. Hence, the size of the action space, similar to the DNN's size, does not depend on the number of UEs.
In conclusion, the size of the whole training system depends only on the number of IAB-nodes, which is small in practice. Therefore, we can assume our proposed approach to be scalable.

In the policy execution phase, the agents independently apply the individual policies without any need to communicate with the central entity or consider other agents' actions. An agent simply infers its optimal action from its local observation. Therefore, the computational complexity of the proposed MARL-based approach during its execution phase is determined only by the complexity of each local policy.


\end{document}